\documentclass[10pt,twocolumn,showpacs,amsmath,amssymb,floatfix,superscriptaddress]{revtex4-2}

\usepackage{graphicx}
\usepackage{dcolumn}
\usepackage{bm}
\usepackage{xcolor}
\usepackage{txfonts}
\usepackage{microtype}
\usepackage{epsfig}
\usepackage{hyperref}

\begin{document}

\title{Generation of  ultrabrilliant polarized attosecond electron bunches via dual-wake injection}
\author{Ting Sun}
\affiliation{Ministry of Education Key Laboratory for Nonequilibrium Synthesis and Modulation of Condensed Matter, Shaanxi Province Key Laboratory of Quantum Information and Quantum Optoelectronic Devices, School of Physics, Xi'an Jiaotong University, Xi'an 710049, China}
\author{Qian Zhao}\email{zhaoq2019@xjtu.edu.cn}
\affiliation{Ministry of Education Key Laboratory for Nonequilibrium Synthesis and Modulation of Condensed Matter, Shaanxi Province Key Laboratory of Quantum Information and Quantum Optoelectronic Devices, School of Physics, Xi'an Jiaotong University, Xi'an 710049, China}
\author{Feng Wan}
\affiliation{Ministry of Education Key Laboratory for Nonequilibrium Synthesis and Modulation of Condensed Matter, Shaanxi Province Key Laboratory of Quantum Information and Quantum Optoelectronic Devices, School of Physics, Xi'an Jiaotong University, Xi'an 710049, China}
\author{Yousef I. Salamin}
\affiliation{Department of Physics, American University of Sharjah, Sharjah, POB 26666 Sharjah,  United Arab Emirates}
\author{Jian-Xing Li}\email{jianxing@xjtu.edu.cn}
\affiliation{Ministry of Education Key Laboratory for Nonequilibrium Synthesis and Modulation of Condensed Matter, Shaanxi Province Key Laboratory of Quantum Information and Quantum Optoelectronic Devices, School of Physics, Xi'an Jiaotong University, Xi'an 710049, China}

\date{\today}
	
\begin{abstract}
Laser wakefield acceleration is paving the way for the next generation of electron accelerators, for their own sake and as radiation sources. A controllable dual-wake injection scheme is put forward here to generate an ultrashort triplet electron bunch with high brightness and high polarization, employing a radially polarized laser as a driver. We find that the dual wakes can be driven by both transverse and longitudinal components of the laser field in the quasi-blowout regime, sustaining the laser-modulated wakefield which facilitates the sub-cycle and transversely-split injection of the  triplet bunch. Polarization of the triplet bunch can be highly preserved due to the laser-assisted collective spin precession and the non-canceled transverse spins. In our three-dimensional particle-in-cell simulations, the triplet electron bunch, with duration about $500$ as, six-dimensional brightness exceeding $10^{14}$ A/m$^2$/0.1$\%$  and polarization over $80\%$, can be generated using a few-terawatt laser. Such an electron bunch could play an essential role in many applications, such as ultrafast imaging, nuclear structure and high-energy physics studies, and the operation of coherent radiation sources.
\end{abstract}

\maketitle
Laser wakefield acceleration (LWFA), especially in the blowout regime driven by an intense laser pulse, can accelerate electrons within the focusing and accelerating phases of a plasma wave (essentially in one quarter of a plasma wavelength) and can generate high-brightness femtosecond electron bunches \cite{tajima1979,lu2006,faure2006,lundh2011,esarey2009}. Quality of the wakefield electron bunch is strongly injection-scheme dependent, and the controllable schemes include field-ionization injection \cite{chen2006,pak2010,McGuffey2010,chen2012}, density-gradient injection \cite{geddes2008,gonsalves2011} and magnetic-field controlled injection \cite{vieira2011magnetic,zhao2018}. Highly localized injection can be implemented in order to accelerate ultrashort electron bunches even at the attosecond scale in LWFA \cite{luttikhof2010,tooley2017,zhao2019,kim2021}. A femtosecond (or attosecond) electron bunch with very high brightness underpins ultrafast imaging and microscopy \cite{baum2007,krausz2009,corde2013,miller2014,morimoto2018} and can be used as an injection beam in a multi-stage wakefield acceleration setup to achieve high-quality GeV particle energies \cite{luo2018,zhu2023}. Six-dimensional (6D) brightness, which characterizes the high-quality wakefield electron bunch \cite{dimitri2014}, can reach about $10^{15}$ A/m$^2$/0.1$\%$ in experiments \cite{wang2016}, and is utilized in the operation of a free electron laser for injection \cite{huang2012,wang2021,pompili2022,labat2023}, and in the generation of compact monoenergetic sources of $\gamma$-rays and coherent X-rays \cite{taphuoc2012,powers2014,ma2023compact,brummer2022}.
It has also been recently proposed that the polarized electrons can be generated in a controllable LWFA scheme \cite{vieira2011polarized,wen2019prl,wu2019pre,wu2019njp,nie2021,fan2022}, in which a prepolarized gas target can be prepared experimentally, with densities reaching $10^{19}-10^{20}$ cm$^{-3}$, via ultraviolet photodissociation of a hydrogen-halogen molecule \cite{rakitzis2003,sofikitis2008,sofikitis2018,spiliotis2021}.
The polarization can enhance the capacity of a high-brightness electron bunch for utilization in a range of applications \cite{sun2022}, such as investigating the properties of a magnetic material \cite{gay2009}, nuclear structure studies \cite{abe1995,alexakhin2007} and the construction of polarized $\gamma$-ray sources \cite{li2020,wang2022,ma2023compact}.

For production of a highly polarized electron bunch in the plasma wakefield, depolarization must be mitigated. The characteristic depolarization in a wakefield originates predominantly from the sheath-current-induced azimuthal magnetic field during injection \cite{wen2019prl,fan2022}. Specifically, depolarization results from two sources: (i) continuous injection of the electrons induces disperse precessions of initial spins due to the evolving wakefield, and (ii) the azimuthally symmetric distribution of  transverse spins results in cancelation. Although depolarization from source (i) can be mitigated by fine control of the driver and plasma parameters for the moderately relativistic intensity of wakefield, the attainable current intensity and brightness are significantly limited \cite{wen2019prl,wu2019pre,wu2019njp,nie2021,barth2013}. Because of the azimuthal symmetry of the sheath current, depolarization from source (ii) can not be eliminated in the currently available injection schemes, to the best of our knowledge. In principle, spin precession of the sheath electrons in the blowout regime can be synchronized via ultrashort injection, even at the attosecond scale, which results in the collective spin rotation of an electron bunch, thus circumventing depolarization from source (i). Nevertheless, the currently available methods of attosecond injection in the blowout regime still cause depolarization from source (ii) \cite{luttikhof2010,tooley2017,zhao2019,kim2021}.
Still, generation of a high-brightness polarized electron bunch remains challenging.

\begin{figure}[!t]	
\setlength{\abovecaptionskip}{0.2cm}  	
\centering\includegraphics[width=0.9\linewidth]{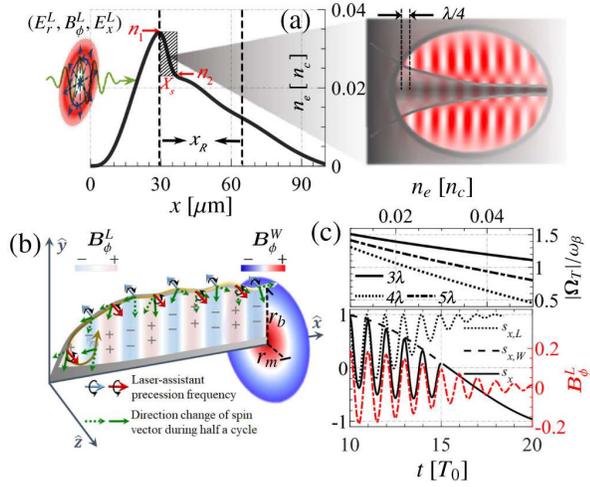}
\caption{(a) Plasma density profile with a transition length $X_s$ between $n_1$ and $n_2$, with the Rayleigh length $x_R$ of a RPL pulse also labeled. The schematic dual-wake plasma bubble is overlapped by the analytical laser intensity $(E_x^L)^2+(E_r^L)^2$ used in the simulations.
(b) Sketch of the spin precession of an electron inside a plasma bubble. (c) Top: Ratio of spin precision frequency $\Omega_T$ and betatron frequency $\omega_\beta$ vs. $n_e$, for different laser spot sizes. Bottom: Spin precession vs. time in laser field ($s_{x,L}$), wakefield ($s_{x,W}$) and superposed field ($s_{x}$).
}
\label{model}
\end{figure}
\begin{figure}[!t]	
\setlength{\abovecaptionskip}{0.2cm}  	
\centering\includegraphics[width=0.9\linewidth]{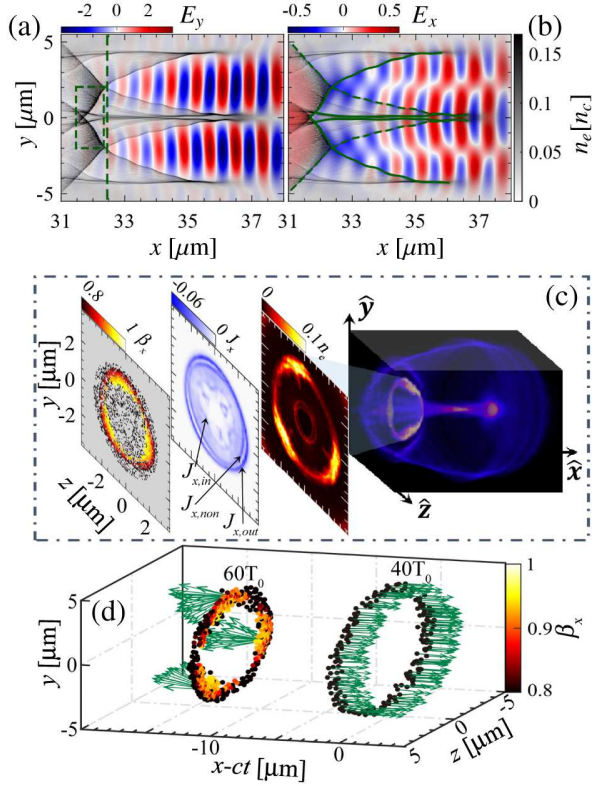}
	\caption{Snapshot at a propagation time $t=52T_0$ of the plasma density $n_e$ (gray) with (a) the transverse electric fields $E_y$ and (b) the longitudinal electric field $E_x$, superimposed upon it. The injection region is indicated within a dashed rectangle and structure of the wakes is indicated by solid and dashed lines. (c) 3D plasma density with the projection of injected electrons on the $y-z$ plane. Transverse distributions of normalized current density $J_x$ and longitudinal velocity $\beta_x$ are sliced at the dash-lined position of (a). (d) Sheath electrons with the spin directions (green arrows) before and after injection.
}
\label{snapshot}
\end{figure}
\begin{figure}[!t]	
\setlength{\abovecaptionskip}{0.2cm}  	
\centering\includegraphics[width=0.9\linewidth]{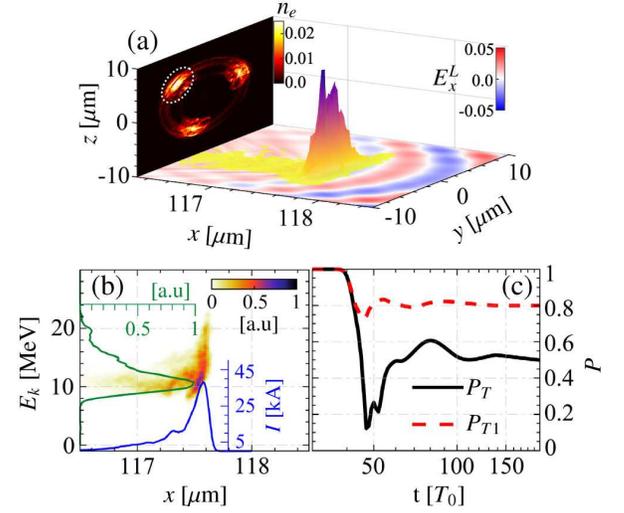}
	\caption{ (a) Transverse and longitudinal density distributions of injected electron bunches at a propagation time $t=160 T_0$, and the laser's longitudinal electric field $E_x^L$ is mapped on the propagation plane.
(b) Phase-space distribution of one beam of the triplet bunch [shown inside a dotted ellipse in (a)] with the energy spectrum and current distribution. (c) Temporal evolution of polarization for the triplet bunch ($P_{T}$) and one of its component beams ($P_{T1}$).}
\label{beam}
\end{figure}
\begin{figure}[!t]	
\setlength{\abovecaptionskip}{0.2cm}  	
\centering\includegraphics[width=0.9\linewidth]{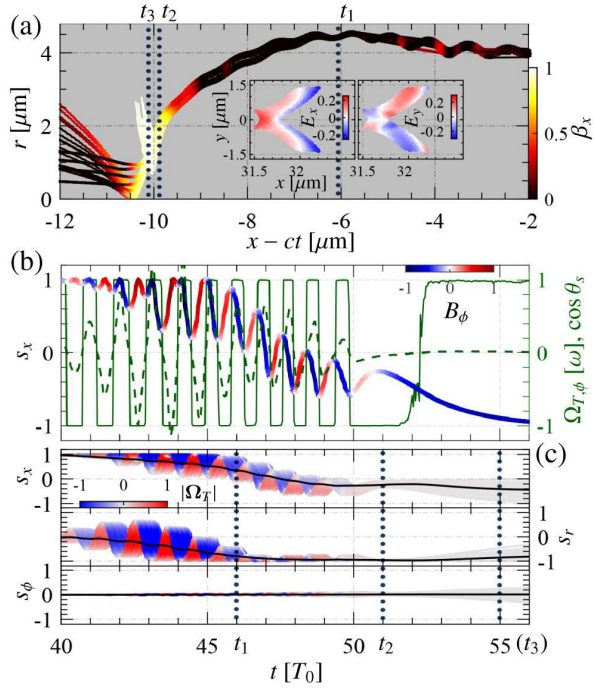}
\caption{
	 (a) Trajectories of the randomly selected sheath electrons from the injection region in Fig. \ref{snapshot}(a), color-coded by the longitudinal velocities. Insets: Snapshot of the tracked electrons at $52T_0$, color-coded by the sensed longitudinal and transverse fields.
	 (b) Left axis: Precession of $s_x$ of an injected electron color-coded by its transient magnetic field $B_\phi$. Right axis: Temporal evolution of $\cos{\theta_s}$ (green-solid line) and $\Omega_{T,\phi}$ (green-dashed line).
	 (c) Precession of  the spin components of electrons in one beam of the triplet bunch, color-coded by the transient precession frequency. The averaged spin components are represented by black-solid lines.}
\label{physics}
\end{figure}
\begin{figure}[!t]	
\setlength{\abovecaptionskip}{0.2cm}  	
\centering\includegraphics[width=0.9\linewidth]{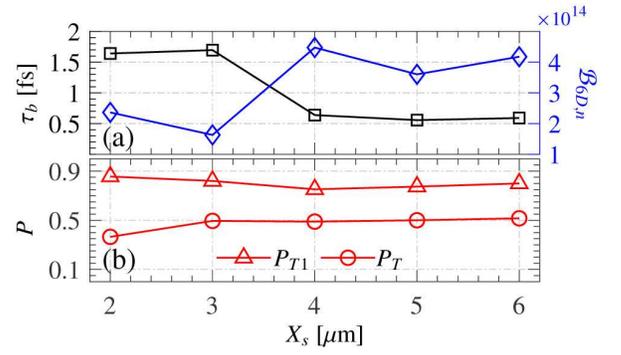}
	\caption{(a) Bunch duration $\tau_b$ and 6D brightness $\mathcal{B}_{6D,n}$ of one beam of the triplet bunch vs. the scale length $X_s$ of the density jump. (b) Variations of $P_{T}$ and $P_{T1}$ with $X_s$.
}
\label{paras}
\end{figure}

In this Letter, we put forward a dual-wake injection scheme for LWFA, as a result of which an ultrashort electron bunch with high brightness and high polarization can be generated. In this scheme, a tightly focused radially polarized laser (RPL) pulse propagates through a plasma with the density profile of a near-critical-density peak and a sharp downward density jump, driving dual wakes in the quasi-blowout wake (QBW) regime \cite{shaw2017}; see Fig. \ref{model}(a). The laser-modulated wakefield in the dual wakes induces oscillatory spin precession of the sheath electrons, whereby the initial spin directions can be reversed at the end of the injection stage; see Fig. \ref{model} (b). The principal parameters of laser and plasma density, modelling the laser-assisted spin precession, are elucidated analytically; see Fig. \ref{model} (c).
Three-dimensional (3D) Particle-in-Cell (PIC) simulations indicate that the driven dual wakes cause the sub-cycle injection of the transversely-split triplet bunch, and demonstrate that the depolarization is significantly mitigated due to the approximately collective spin precession and transverse spins that survive cancellation; see Figs. \ref{snapshot} and \ref{beam}. The dynamical processes associated with the spin precessions are elucidated by carefully tracking the trajectories of individual electrons in one beam of the triplet bunch; see Fig. \ref{physics}.
 Robust generation of ultrashort electron bunches with high brightness and high polarization is tested by varying the transition length of density jump; see Fig. \ref{paras}.

The blowout regime arises when the laser pulse duration $\tau_L$ matches the plasma wave period, i.e., when $c\tau_L\lesssim2\sqrt{a_0}/k_p$  \cite{esarey2009}. Here $k_p=\omega_p/c$, with the plasma frequency $\omega_p$   and the speed of light in vacuum $c$, and $a_0$ is amplitude of the laser vector potential. Violation of this condition for a relatively long $\tau_L$ results in the QBW regime in which the laser pulse overlaps the wake bucket. A tightly focused RPL pulse has an intense longitudinal electric field $E_x^L$, a radial electric field $E_r^L$ and an azimuthal magnetic field $B_\phi^L$ \cite{salamin2006njp,varin2013}, resembling the configuration of the plasma wakefield and, thus, can be used to excite the laser-modulated wakefield in the QBW regime. For a strongly nonlinear plasma wake, the betatron frequency, $\omega_\beta=\omega_p/\sqrt{2\gamma_e}$, with $\gamma_e\simeq\sqrt{1+a_0^2/2}$ in the ponderomotive approximation, characterizes the formation time of a plasma bubble by half the betatron period $\tau_\beta=2\pi/\omega_\beta$ \cite{tsakiris2000}.
A RPL pulse of wavelength $\lambda=0.8$ $\mu$m, power 10 TW, pulse duration $5T_0$ and a spot size of $w_0=4\lambda$, has $\tau_\beta/2$ in the range $\simeq8T_0$ to $5T_0$, which corresponds to a density range of $n_e=0.02n_c$ to $0.05n_c$. Here, $T_0=\lambda/c$, and $n_c$ is the critical density.

In a feasible setup, the plasma density profile can be initialized with a near-critical density peak and  a sharp downward density jump to realize the controllable injection of high-quality electron bunches, which can be prepared experimentally by triggering a shock front in a supersonic gas jet \cite{schmid2010,buck2013,chou2016}.  This density profile can be modeled by an analytical function with the combination of a Gamma distribution function and an asymmetric double sigmodal function \cite{supplement}, where the density jump can be characterized by the transition length $X_s$ and the density ratio $n_2/n_1$ and parameterized according to the kinetic theory of gases \cite{mott1951}; see left panel of Fig. \ref{model}(a).
Thus, under the conditions of QBW regime and tight focusing of RPL pulse, the dual wakes can be driven in this density profile with a $E_r^L$-driven donut wake and a $E_x^L$-driven non-closed wake; see right panel of Fig. \ref{model}(a). The sheath electrons ride on the laser phases in the injection region of the dual wakes and get modulated by both the longitudinal and transverse laser fields, with a $\lambda/4$ overlap. The longitudinal trapping condition in the RPL-driven QBW is modified due to the superposed laser vector potential and described by variation of the pseudopotential $\Delta \psi \leq \sqrt{1+\beta_\perp^2}/\gamma_b-1 + \Delta \psi^{\rm{L}}$, where $\Delta\psi^{L}=ev_b/m_ec^2[A_x^L(r_f)-A_x^L(r_i)]$, in which $E_x^L=-\partial A_x^L/\partial t$, and $r_i$ and $r_f$, respectively, are the initial and final radial positions of a sheath electron \cite{zhao2018}. Here $m_e$ and $e$ are the electron mass and charge, respectively. A sheath electron should be injected into the acceleration phase of $E_x^L$ in order for it to copropagate with the laser, which implies that the sheath electrons experience integer numbers of laser cycles between $A_x^L(r_f)$ and $A_x^L(r_i)$, and undergo sub-cycle injection.

In what follows, $a_l=\omega/\omega_p$, unless explicitly stated, and the electric and magnetic fields are normalized by $e/(m_ec\omega)$ and $e/(m_e\omega)$, respectively, with $\omega$ the laser frequency. By analogy to the donut wake driven by a Laguerre-Gaussian laser \cite{vieira2014}, components of the spherical bubble field are $E_r^W=k_p(r-r_m)/4a_l$, $B_\phi^W=-k_p(r-r_m)/4a_l$, and $E_x^W=k_p(x-r_b-v_bt)/2a_l$, where $r_m\simeq w_0/\sqrt{2}$, $r_b\simeq2\sqrt{a_\perp}$, with a transverse laser vector potential $a_\perp$ and a bubble velocity $v_b$. For a bubble sheath electron with spin vector $\bm{s}=(s_r,s_\phi,s_x)$ in cylindrical coordinates, irrespective of the Sokolov-Ternov effect ($\sim$ms scale in our model) \cite{supplement} and the negligible Stern–Gerlach force in LWFA \cite{thomas2020}, the spin precession can be described by the Thomas-Bargmann-Michel-Telegdi (T-BMT) equation \cite{mane2005} $d\bm{s}/dt=\bm{\Omega}\times\bm{s}=(\bm{\Omega}_T+\bm{\Omega}_a)\times\bm{s}$, where $\bm{\Omega}_a=a\left[\bm{B}-\frac{\gamma_e}{\gamma_e+1}\bm{\beta}(\bm{\beta}\times\bm{B})-\bm{\beta}\times\bm{E}\right]$, with the anomalous magnetic moment of the electron $a\approx1.16\times10^{-3}$,
 and $\bm{\Omega}_T=\frac{\bm{B}}{\gamma_e}-\frac{1}{(\gamma_e+1)}\bm{\beta}\times\bm{E}$.
Recall that $|\bm{\Omega}_a|\ll|\bm{\Omega}_T|$, $\bm{\Omega}_T^I\simeq B_\phi^W\hat{\bm\phi}/\gamma_e$ dominates the spin precession of a wakefield electron during the injection stage, and $\bm{\Omega}_T^{II}\simeq(B_\phi^W-E_r^W)\hat{\bm{\phi}}/\gamma_e=-F_{W,\perp}\hat{\bm{\phi}}/\gamma_e$ dominates during the acceleration stage \cite{vieira2011polarized,wen2019prl,fan2022}.
In the laser-modulated wakefield, the spin precession of a sheath electron is also influenced by $B_\phi^L$ and displays oscillatory precession in each optical cycle. Qualitatively, $B_\phi^W$ causes the net precession and $B_\phi^L$ causes the change of precession direction at the end of the injection stage, resulting in the laser-assisted spin precession of a sheath electron and reversal of the spin vector; see in Fig. \ref{model}(b).

The laser-assisted spin precession can be elucidated by resorting to the T-BMT equation, using the spherical bubble fields.
Variation of the ratio $|\bm{\Omega}_T|/\omega_\beta$ with the plasma density, for different values of $w_0$, is shown in Fig. \ref{model}(c)-top. Note that $|\bm{\Omega}_T|$ can be comparable to the betatron frequency $\omega_\beta$ in the bubble wakefield. The critical $|\bm{\Omega}_T|/\omega_\beta=1$ means that the spin precession and betatron oscillation are synchronous, causing $\bm{s}$ to reverse direction as a sheath electron slips back to the bottom of the bubble, and the depolarization increases linearly with density. To distinguish this from the case of the wakefield \cite{vieira2011polarized,wen2019prl,fan2022}, the laser-assisted  precession frequencies are denoted by $\bm{\Omega}_T^{'I}\simeq(B_\phi^L+B_\phi^W)\hat{\bm{\phi}}/\gamma_e$ and $\bm{\Omega}_T^{'II}\simeq(F_{W,\perp}+F_{L,\perp})\hat{\bm{\phi}}/\gamma_e$, corresponding to the injection and acceleration stages, respectively, of the laser-modulated wakefield, with $F_{L,\perp}=E_r^L-B_\phi^L$ resulting from employing the RPL fields beyond the paraxial approximation. Calculations, employing $s_x=\cos{(|\bm{\Omega}_T^{'I}|\tau_\beta/2)}$, indicate that the net precession, from $s_x=1$ to $s_x\simeq0$ at $15T_0$, may be obtained from $\bm{B}_\phi^{W}$, whereas the upward precession direction (from $s_x=-1$ to $s_x=1$ and vice versa) is due to the focused phase of $\bm{B}_\phi^{L}$; see Fig. \ref{model}(c)-bottom. After $t=15T_0$, $s_x$ precesses at $\bm{\Omega}_T^{'II}$ with a downward precession direction due to the laser-phase-locked injection into the defocused field with negative $F_{L,\perp}+F_{W,\perp}$.


In order to visualize the polarized acceleration in the dual-wake injection scheme, 3D PIC simulations have been carried out using EPOCH and incorporating the T-BMT equations \cite{arber2015,wan2022}. In each case, a moving window along the $x-$direction is used, with a $25\lambda(x)\times25\lambda(y)\times25\lambda(z)$ simulation box divided into $640\times160\times160$ cells.
The explicit truncated series expressions modelling the fields of a focused RPL pulse are used\cite{salamin2006njp}, initialized with 10 TW power, $w_0=3$ $\mu$m and $\tau_L=10$ fs. With $X_s=6$ $\mu$m and $n_2/n_1=0.66$, the dual wakes sustaining the laser-modulated wakefield are driven near the focal plane (at $x=30$ $\mu$m); see Figs. \ref{snapshot}(a) and (b).  $F_{L,\perp}$ drives the donut wake bucket (solid lines) with closed outer-sheath density and almost merged inner-sheath density, and the intense $E_x^L$ sweeps over the inner-sheath density and drives the outward non-closed sheath density  (solid-dashed lines) due to the near-axis defocusing field  \cite{supplement}.
The annular sheath electrons are accelerated and split into the Clover-like distribution transversely due to the transversely-structured current density; see Fig. \ref{snapshot}(c). The inner-sheath current $J_{x,in}$ senses the transversely non-uniform longitudinal field $E_x$ at the bottom of the donut wake [Fig. \ref{snapshot}(b)], leading to the deformed transverse distribution of the non-closed sheath current $J_{x,non}$.  The current $J_{x,non}$ intersects with the outer-sheath current $J_{x,out}$, which weakens $J_x$ in the crossing sectors. As a result,  pseudopotential $\psi$ of the wakefield exhibits transverse asymmetry, according to Posisson's equation $-\nabla_\perp^2\psi=(\rho-J_x)$, with a transversely-structured current source $J_x$, which gives rise to (the transversely-asymmetric) $E_x^W$ \cite{lu2006,supplement}. Consequently, the Clover-like sheath electrons are injected with the sub-cycle duration and transversely-split beams, which constitutes the ``dual-wake injection''. Furthermore, because of the approximate match between $\omega_\beta$ and $\Omega_T^{'I}$, spins of the injected electrons possess a collective precession orientation toward $s_x=-1$; see Fig. \ref{snapshot}(d). Statistical polarization of the electrons is $P=\sqrt{\bar{s}_x^2+\bar{s}_y^2+\bar{s}_z^2}$, with $\bar{s}_{x,y,z}$ the components averaged over the particle number.

The injected Clover-like sheath electrons undergo direct laser acceleration by $E_x^L$ and get transversely modulated into the three beams making up the ``triplet bunch''; see Fig. \ref{beam}(a). Because of the laser spot defocusing beyond the focal plane, an opening angle $\delta\theta\simeq0.135$ rad is associated with the triplet bunch, leading to the completely isolated beams of that bunch (see details in Fig. 3 of  \cite{supplement}).
The corresponding phase-space distribution of one beam of the triplet bunch is shown in Fig. \ref{beam} (b). The quasi-monoenergetic attosecond bunch has a pulse duration of $\sim T_0/4$, a normalized transverse emittance of $\varepsilon_{n,y(z)}\simeq0.02(0.03)$ mm$\cdot$mrad,  and a peak current of $\sim38$ kA.  Evolution of polarization of the triplet bunch is distinguished from that of one of its component beams; see Fig. \ref{beam}(c). $P_{T}$ comes essentially from the averaged longitudinal spin component $\bar{s}_x$, due to cancellation of the transverse spin component $s_r$.  $P_{T}$ drops to zero, followed by a one-period oscillation, before  rising to the final stable $50\%$, during which spin vectors of the injected electrons successively undergo full reversal within a time of about $\tau_\beta/2$; as explained in the caption of Fig. \ref{physics}. On the other hand, $P_{T1}$ is highly preserved during the entire evolution, due to the approximately collective precession with the non-vanishing polarization from $s_r$.

The sub-cycle injection and laser-assisted spin precession in the laser-modulated wakefield can both be clarified by tracking the trajectories of sheath electrons.
The fields sensed by the electrons in the injection region exhibit overlap of $\lambda/2$ focusing phase and $\lambda/4$ acceleration phase; see the insets of Fig. \ref{physics}(a).
The corresponding trajectories indicate that the sheath electrons slip to the center of the wake bucket at $t_1$, and the injection occurs between $t_2$ and $t_3$. During the injection, only the sheath electrons located by laser acceleration phase are injected and get modulated into an attosecond bunch with $\sim T_0/4$ duration; see trajectories after $t_3$ in Fig. \ref{physics}(a).
The injected electrons maintain polarization during the injection stage. Direction of $\bm{\Omega}_T(t)$ will be expressed in terms of $\cos{\theta_s}$, where $\theta_s$ is the angle it makes with the initial $\bm{\Omega}_T(t_0)$. In addition, $|\bm{\Omega}_T|$ approximately equals its azimuthal component $\Omega_{T,\phi}$ by recalling the azimuthal vectors $\bm{\Omega}_T^{'I}$ and $\bm{\Omega}_T^{'II}$ (see details in \cite{supplement}). Temporal evolution of the spin-precession frequency, shown in Fig. \ref{physics}(b), has been obtained from investigating the dynamics of a randomly chosen electron in the triplet bunch. The square-wave-like variation of $\cos{\theta_s}$ is consistent with the oscillating $\Omega_{T,\phi}$, which confirms the azimuthal nature of $\bm{\Omega}_T$. Beginning with the symmetric oscillation before $t_1$, $\Omega_{T,\phi}$ exhibits asymmetric oscillations between $t_1$ and $t_2$, due to the rising wakefield $B_\phi^W$, which leads to the net precession from $s_x=1$ to $s_x\simeq0$ within $5T_0$. As explained, in principle, in Fig. \ref{model}(c), the precession direction of the electron is upward before $t_2$ and is reversed again towards $s_x=-1$ after the injection,  resulting in high polarization. Because of the similar $|\bm{\Omega}_T|$ for each electron during the injection stage, $s_x$ and $s_r$ of them possess almost the same precession before $t_2$, which leads to the collective rotation of spin vectors and the non-deteriorating $P_{T1}$; see Fig. \ref{physics}(c). Planar motion of the sheath electrons indicates that there is almost no precession along the azimuthal direction before injection, while the precession of $s_\phi$ occurs during laser modulation of the electrons after injection.

Robustness of the dual-wake injection scheme can be examined by changing parameters of the sharp downward density jump. With the theoretical estimates (see details in \cite{supplement}), $X_s$ and $n_2/n_1$ can, respectively, be approximated as (2, 3, 4, 5, 6) $\mu$m, and (0.4, 0.51, 0.56, 0.6, 0.66), for the sake of parameterizing the density profile in Fig. \ref{model}(a). For values of $X_s$ smaller than the plasma wavelength $\lambda_{p1}$ ($\simeq4.3$ $\mu$m at $n_1$), the shock-front injection, which leads to injection of the density peak of the wake bucket driven at $n_1$ into the adjacent wake bucket driven at $n_2$ \cite{buck2013}, leads to the injected bunch with length of $\sim\lambda/2$. As $X_s$ increases beyond $\lambda_{p1}$, the relatively gentle density gradient generates the attosecond bunch of length $\lambda/4$; see Fig. \ref{paras}(a).
The 6D brightness $\mathcal{B}_{6D,n}$ increases up to $\sim4\times10^{14}$ A/m$^2$/0.1$\%$ as $X_s>3$ $\mu$m where the density-gradient injection leads to the narrower energy spectra \cite{supplement}. This is close to the maximum $\sim10^{15}$ A/m$^2$/0.1$\%$, obtained by state-of-the-art LWFA and linac bunches \cite{dimitri2014,wang2016}. $P_{T}$ and $P_{T1}$ are almost independent of $X_s$, because spin precession of the sheath electrons mainly occurs during the wake formation time $\tau_\beta/2$, which undergoes minor changes due to slight variations in $X_s$; see Fig. \ref{paras}(b).

With currently available laser technology, a femtosecond RPL pulse with peak power exceeding 85 GW is attainable in experiments \cite{carbajo2014,kong2019}. Terawatt-power beams seem feasible via the technology of optical parametric amplification \cite{zhong2021}. So, the potential exists for such a system to be used to experimentally realize our dual-wake injection scheme in a RPL-driven QBW regime. 

In conclusion, generation of the ultrashort triplet electron bunch with high brightness and high polarization  has been investigated in the dual-wake injection scheme for LWFA. In this scheme, the laser-modulated wakefiled in a donut QBW leads to the sub-cycle and the transversely-split injection of the triplet electron bunch.
High polarization of the triplet electron bunch can be preserved due to the laser-assisted spin precession and non-canceled transverse spins. Our 3D PIC simulations demonstrate robust generation of the ultrashort electron bunches with brightness over $10^{14}$ A/m$^2$/0.1$\%$ and polarization exceeding $80\%$. Such an electron bunch is advantageous as a source of polarized $\gamma$-rays and  in applications in ultrafast imaging, nuclear physics and high-energy physics studies.

{\it Acknowledgement:}
We thank S. Weng and M. Chen for fruitful discussions.
The work is supported by the National Natural Science Foundation of China (Grants No. 12022506, No. U2267204, No. 12275209, No. 12105217), the Foundation of Science and Technology on Plasma Physics Laboratory (No. JCKYS2021212008), the Open Foundation of Key Laboratory of High Power Laser and Physics, Chinese Academy of Sciences (SGKF202101), and the Shaanxi Fundamental Science Research Project for Mathematics and Physics (Grant No. 22JSY014). YIS is supported by an American University of Sharjah Faculty Research Grant (FRG-22).

\bibliography{ref-RPL}

\begin{thebibliography}{68}%
\makeatletter
\providecommand \@ifxundefined [1]{%
 \@ifx{#1\undefined}
}%
\providecommand \@ifnum [1]{%
 \ifnum #1\expandafter \@firstoftwo
 \else \expandafter \@secondoftwo
 \fi
}%
\providecommand \@ifx [1]{%
 \ifx #1\expandafter \@firstoftwo
 \else \expandafter \@secondoftwo
 \fi
}%
\providecommand \natexlab [1]{#1}%
\providecommand \enquote  [1]{``#1''}%
\providecommand \bibnamefont  [1]{#1}%
\providecommand \bibfnamefont [1]{#1}%
\providecommand \citenamefont [1]{#1}%
\providecommand \href@noop [0]{\@secondoftwo}%
\providecommand \href [0]{\begingroup \@sanitize@url \@href}%
\providecommand \@href[1]{\@@startlink{#1}\@@href}%
\providecommand \@@href[1]{\endgroup#1\@@endlink}%
\providecommand \@sanitize@url [0]{\catcode `\\12\catcode `\$12\catcode
  `\&12\catcode `\#12\catcode `\^12\catcode `\_12\catcode `\%12\relax}%
\providecommand \@@startlink[1]{}%
\providecommand \@@endlink[0]{}%
\providecommand \url  [0]{\begingroup\@sanitize@url \@url }%
\providecommand \@url [1]{\endgroup\@href {#1}{\urlprefix }}%
\providecommand \urlprefix  [0]{URL }%
\providecommand \Eprint [0]{\href }%
\providecommand \doibase [0]{https://doi.org/}%
\providecommand \selectlanguage [0]{\@gobble}%
\providecommand \bibinfo  [0]{\@secondoftwo}%
\providecommand \bibfield  [0]{\@secondoftwo}%
\providecommand \translation [1]{[#1]}%
\providecommand \BibitemOpen [0]{}%
\providecommand \bibitemStop [0]{}%
\providecommand \bibitemNoStop [0]{.\EOS\space}%
\providecommand \EOS [0]{\spacefactor3000\relax}%
\providecommand \BibitemShut  [1]{\csname bibitem#1\endcsname}%
\let\auto@bib@innerbib\@empty
\bibitem [{\citenamefont {Tajima}\ and\ \citenamefont
  {Dawson}(1979)}]{tajima1979}%
  \BibitemOpen
  \bibfield  {author} {\bibinfo {author} {\bibfnamefont {T.}~\bibnamefont
  {Tajima}}\ and\ \bibinfo {author} {\bibfnamefont {J.~M.}\ \bibnamefont
  {Dawson}},\ }\bibfield  {title} {\bibinfo {title} {Laser electron
  accelerator},\ }\href {https://doi.org/10.1103/PhysRevLett.43.267} {\bibfield
   {journal} {\bibinfo  {journal} {Phys. Rev. Lett.}\ }\textbf {\bibinfo
  {volume} {43}},\ \bibinfo {pages} {267} (\bibinfo {year} {1979})}\BibitemShut
  {NoStop}%
\bibitem [{\citenamefont {Lu}\ \emph {et~al.}(2006)\citenamefont {Lu},
  \citenamefont {Huang}, \citenamefont {Zhou}, \citenamefont {Mori},\ and\
  \citenamefont {Katsouleas}}]{lu2006}%
  \BibitemOpen
  \bibfield  {author} {\bibinfo {author} {\bibfnamefont {W.}~\bibnamefont
  {Lu}}, \bibinfo {author} {\bibfnamefont {C.}~\bibnamefont {Huang}}, \bibinfo
  {author} {\bibfnamefont {M.}~\bibnamefont {Zhou}}, \bibinfo {author}
  {\bibfnamefont {W.~B.}\ \bibnamefont {Mori}},\ and\ \bibinfo {author}
  {\bibfnamefont {T.}~\bibnamefont {Katsouleas}},\ }\bibfield  {title}
  {\bibinfo {title} {Nonlinear {{Theory}} for {{Relativistic Plasma
  Wakefields}} in the {{Blowout Regime}}},\ }\href
  {https://doi.org/10.1103/PhysRevLett.96.165002} {\bibfield  {journal}
  {\bibinfo  {journal} {Phys. Rev. Lett.}\ }\textbf {\bibinfo {volume} {96}},\
  \bibinfo {pages} {165002} (\bibinfo {year} {2006})}\BibitemShut {NoStop}%
\bibitem [{\citenamefont {Faure}\ \emph {et~al.}(2006)\citenamefont {Faure},
  \citenamefont {Rechatin}, \citenamefont {Norlin}, \citenamefont {Lifschitz},
  \citenamefont {Glinec},\ and\ \citenamefont {Malka}}]{faure2006}%
  \BibitemOpen
  \bibfield  {author} {\bibinfo {author} {\bibfnamefont {J.}~\bibnamefont
  {Faure}}, \bibinfo {author} {\bibfnamefont {C.}~\bibnamefont {Rechatin}},
  \bibinfo {author} {\bibfnamefont {A.}~\bibnamefont {Norlin}}, \bibinfo
  {author} {\bibfnamefont {A.}~\bibnamefont {Lifschitz}}, \bibinfo {author}
  {\bibfnamefont {Y.}~\bibnamefont {Glinec}},\ and\ \bibinfo {author}
  {\bibfnamefont {V.}~\bibnamefont {Malka}},\ }\bibfield  {title} {\bibinfo
  {title} {Controlled injection and acceleration of electrons in plasma
  wakefields by colliding laser pulses},\ }\href
  {https://doi.org/10.1038/nature05393} {\bibfield  {journal} {\bibinfo
  {journal} {Nature}\ }\textbf {\bibinfo {volume} {444}},\ \bibinfo {pages}
  {737} (\bibinfo {year} {2006})}\BibitemShut {NoStop}%
\bibitem [{\citenamefont {Lundh}\ \emph {et~al.}(2011)\citenamefont {Lundh},
  \citenamefont {Lim}, \citenamefont {Rechatin}, \citenamefont {Ammoura},
  \citenamefont {{Ben-Ismail}}, \citenamefont {Davoine}, \citenamefont
  {Gallot}, \citenamefont {Goddet}, \citenamefont {Lefebvre}, \citenamefont
  {Malka},\ and\ \citenamefont {Faure}}]{lundh2011}%
  \BibitemOpen
  \bibfield  {author} {\bibinfo {author} {\bibfnamefont {O.}~\bibnamefont
  {Lundh}}, \bibinfo {author} {\bibfnamefont {J.}~\bibnamefont {Lim}}, \bibinfo
  {author} {\bibfnamefont {C.}~\bibnamefont {Rechatin}}, \bibinfo {author}
  {\bibfnamefont {L.}~\bibnamefont {Ammoura}}, \bibinfo {author} {\bibfnamefont
  {A.}~\bibnamefont {{Ben-Ismail}}}, \bibinfo {author} {\bibfnamefont
  {X.}~\bibnamefont {Davoine}}, \bibinfo {author} {\bibfnamefont
  {G.}~\bibnamefont {Gallot}}, \bibinfo {author} {\bibfnamefont {J.-P.}\
  \bibnamefont {Goddet}}, \bibinfo {author} {\bibfnamefont {E.}~\bibnamefont
  {Lefebvre}}, \bibinfo {author} {\bibfnamefont {V.}~\bibnamefont {Malka}},\
  and\ \bibinfo {author} {\bibfnamefont {J.}~\bibnamefont {Faure}},\ }\bibfield
   {title} {\bibinfo {title} {Few femtosecond, few kiloampere electron bunch
  produced by a laser-plasma accelerator},\ }\href
  {https://doi.org/10.1038/NPHYS1872} {\bibfield  {journal} {\bibinfo
  {journal} {Nat. Phys.}\ }\textbf {\bibinfo {volume} {7}},\ \bibinfo {pages}
  {219} (\bibinfo {year} {2011})}\BibitemShut {NoStop}%
\bibitem [{\citenamefont {Esarey}\ \emph {et~al.}(2009)\citenamefont {Esarey},
  \citenamefont {Schroeder},\ and\ \citenamefont {Leemans}}]{esarey2009}%
  \BibitemOpen
  \bibfield  {author} {\bibinfo {author} {\bibfnamefont {E.}~\bibnamefont
  {Esarey}}, \bibinfo {author} {\bibfnamefont {C.~B.}\ \bibnamefont
  {Schroeder}},\ and\ \bibinfo {author} {\bibfnamefont {W.~P.}\ \bibnamefont
  {Leemans}},\ }\bibfield  {title} {\bibinfo {title} {Physics of laser-driven
  plasma-based electron accelerators},\ }\href
  {https://doi.org/10.1103/RevModPhys.81.1229} {\bibfield  {journal} {\bibinfo
  {journal} {Rev. Mod. Phys.}\ }\textbf {\bibinfo {volume} {81}},\ \bibinfo
  {pages} {1229} (\bibinfo {year} {2009})}\BibitemShut {NoStop}%
\bibitem [{\citenamefont {Chen}\ \emph {et~al.}(2006)\citenamefont {Chen},
  \citenamefont {Sheng}, \citenamefont {Ma},\ and\ \citenamefont
  {Zhang}}]{chen2006}%
  \BibitemOpen
  \bibfield  {author} {\bibinfo {author} {\bibfnamefont {M.}~\bibnamefont
  {Chen}}, \bibinfo {author} {\bibfnamefont {Z.~M.}\ \bibnamefont {Sheng}},
  \bibinfo {author} {\bibfnamefont {Y.~Y.}\ \bibnamefont {Ma}},\ and\ \bibinfo
  {author} {\bibfnamefont {J.}~\bibnamefont {Zhang}},\ }\bibfield  {title}
  {\bibinfo {title} {Electron injection and trapping in a laser wakefield by
  field ionization to high-charge states of gases},\ }\href
  {https://doi.org/10.1063/1.2179194} {\bibfield  {journal} {\bibinfo
  {journal} {J. Appl. Phys.}\ }\textbf {\bibinfo {volume} {99}},\ \bibinfo
  {pages} {056109} (\bibinfo {year} {2006})}\BibitemShut {NoStop}%
\bibitem [{\citenamefont {Pak}\ \emph {et~al.}(2010)\citenamefont {Pak},
  \citenamefont {Marsh}, \citenamefont {Martins}, \citenamefont {Lu},
  \citenamefont {Mori},\ and\ \citenamefont {Joshi}}]{pak2010}%
  \BibitemOpen
  \bibfield  {author} {\bibinfo {author} {\bibfnamefont {A.}~\bibnamefont
  {Pak}}, \bibinfo {author} {\bibfnamefont {K.~A.}\ \bibnamefont {Marsh}},
  \bibinfo {author} {\bibfnamefont {S.~F.}\ \bibnamefont {Martins}}, \bibinfo
  {author} {\bibfnamefont {W.}~\bibnamefont {Lu}}, \bibinfo {author}
  {\bibfnamefont {W.~B.}\ \bibnamefont {Mori}},\ and\ \bibinfo {author}
  {\bibfnamefont {C.}~\bibnamefont {Joshi}},\ }\bibfield  {title} {\bibinfo
  {title} {Injection and trapping of tunnel-ionized electrons into
  laser-produced wakes},\ }\href
  {https://doi.org/10.1103/PhysRevLett.104.025003} {\bibfield  {journal}
  {\bibinfo  {journal} {Phys. Rev. Lett.}\ }\textbf {\bibinfo {volume} {104}},\
  \bibinfo {pages} {025003} (\bibinfo {year} {2010})}\BibitemShut {NoStop}%
\bibitem [{\citenamefont {McGuffey}\ \emph {et~al.}(2010)\citenamefont
  {McGuffey}, \citenamefont {Thomas}, \citenamefont {Schumaker}, \citenamefont
  {Matsuoka}, \citenamefont {Chvykov}, \citenamefont {Dollar}, \citenamefont
  {Kalintchenko}, \citenamefont {Yanovsky}, \citenamefont {Maksimchuk},
  \citenamefont {Krushelnick}, \citenamefont {Bychenkov}, \citenamefont
  {Glazyrin},\ and\ \citenamefont {Karpeev}}]{McGuffey2010}%
  \BibitemOpen
  \bibfield  {author} {\bibinfo {author} {\bibfnamefont {C.}~\bibnamefont
  {McGuffey}}, \bibinfo {author} {\bibfnamefont {A.~G.~R.}\ \bibnamefont
  {Thomas}}, \bibinfo {author} {\bibfnamefont {W.}~\bibnamefont {Schumaker}},
  \bibinfo {author} {\bibfnamefont {T.}~\bibnamefont {Matsuoka}}, \bibinfo
  {author} {\bibfnamefont {V.}~\bibnamefont {Chvykov}}, \bibinfo {author}
  {\bibfnamefont {F.~J.}\ \bibnamefont {Dollar}}, \bibinfo {author}
  {\bibfnamefont {G.}~\bibnamefont {Kalintchenko}}, \bibinfo {author}
  {\bibfnamefont {V.}~\bibnamefont {Yanovsky}}, \bibinfo {author}
  {\bibfnamefont {A.}~\bibnamefont {Maksimchuk}}, \bibinfo {author}
  {\bibfnamefont {K.}~\bibnamefont {Krushelnick}}, \bibinfo {author}
  {\bibfnamefont {V.~Y.}\ \bibnamefont {Bychenkov}}, \bibinfo {author}
  {\bibfnamefont {I.~V.}\ \bibnamefont {Glazyrin}},\ and\ \bibinfo {author}
  {\bibfnamefont {A.~V.}\ \bibnamefont {Karpeev}},\ }\bibfield  {title}
  {\bibinfo {title} {Ionization induced trapping in a laser wakefield
  accelerator},\ }\href {https://doi.org/10.1103/PhysRevLett.104.025004}
  {\bibfield  {journal} {\bibinfo  {journal} {Phys. Rev. Lett.}\ }\textbf
  {\bibinfo {volume} {104}},\ \bibinfo {pages} {025004} (\bibinfo {year}
  {2010})}\BibitemShut {NoStop}%
\bibitem [{\citenamefont {Chen}\ \emph {et~al.}(2012)\citenamefont {Chen},
  \citenamefont {Esarey}, \citenamefont {Schroeder}, \citenamefont {Geddes},\
  and\ \citenamefont {Leemans}}]{chen2012}%
  \BibitemOpen
  \bibfield  {author} {\bibinfo {author} {\bibfnamefont {M.}~\bibnamefont
  {Chen}}, \bibinfo {author} {\bibfnamefont {E.}~\bibnamefont {Esarey}},
  \bibinfo {author} {\bibfnamefont {C.~B.}\ \bibnamefont {Schroeder}}, \bibinfo
  {author} {\bibfnamefont {C.~G.~R.}\ \bibnamefont {Geddes}},\ and\ \bibinfo
  {author} {\bibfnamefont {W.~P.}\ \bibnamefont {Leemans}},\ }\bibfield
  {title} {\bibinfo {title} {Theory of ionization-induced trapping in
  laser-plasma accelerators},\ }\href {https://doi.org/10.1063/1.3689922}
  {\bibfield  {journal} {\bibinfo  {journal} {Phys. Plasmas}\ }\textbf
  {\bibinfo {volume} {19}},\ \bibinfo {pages} {033101} (\bibinfo {year}
  {2012})}\BibitemShut {NoStop}%
\bibitem [{\citenamefont {Geddes}\ \emph {et~al.}(2008)\citenamefont {Geddes},
  \citenamefont {Nakamura}, \citenamefont {Plateau}, \citenamefont {Toth},
  \citenamefont {{Cormier-Michel}}, \citenamefont {Esarey}, \citenamefont
  {Schroeder}, \citenamefont {Cary},\ and\ \citenamefont
  {Leemans}}]{geddes2008}%
  \BibitemOpen
  \bibfield  {author} {\bibinfo {author} {\bibfnamefont {C.~G.~R.}\
  \bibnamefont {Geddes}}, \bibinfo {author} {\bibfnamefont {K.}~\bibnamefont
  {Nakamura}}, \bibinfo {author} {\bibfnamefont {G.~R.}\ \bibnamefont
  {Plateau}}, \bibinfo {author} {\bibfnamefont {C.}~\bibnamefont {Toth}},
  \bibinfo {author} {\bibfnamefont {E.}~\bibnamefont {{Cormier-Michel}}},
  \bibinfo {author} {\bibfnamefont {E.}~\bibnamefont {Esarey}}, \bibinfo
  {author} {\bibfnamefont {C.~B.}\ \bibnamefont {Schroeder}}, \bibinfo {author}
  {\bibfnamefont {J.~R.}\ \bibnamefont {Cary}},\ and\ \bibinfo {author}
  {\bibfnamefont {W.~P.}\ \bibnamefont {Leemans}},\ }\bibfield  {title}
  {\bibinfo {title} {Plasma-density-gradient injection of low
  absolute-momentum-spread electron bunches},\ }\href
  {https://doi.org/10.1103/PhysRevLett.100.215004} {\bibfield  {journal}
  {\bibinfo  {journal} {Phys. Rev. Lett.}\ }\textbf {\bibinfo {volume} {100}},\
  \bibinfo {pages} {215004} (\bibinfo {year} {2008})}\BibitemShut {NoStop}%
\bibitem [{\citenamefont {Gonsalves}\ \emph {et~al.}(2011)\citenamefont
  {Gonsalves}, \citenamefont {Nakamura}, \citenamefont {Lin}, \citenamefont
  {Panasenko}, \citenamefont {Shiraishi}, \citenamefont {Sokollik},
  \citenamefont {Benedetti}, \citenamefont {Schroeder}, \citenamefont {Geddes},
  \citenamefont {{van Tilborg}}, \citenamefont {Osterhoff}, \citenamefont
  {Esarey}, \citenamefont {Toth},\ and\ \citenamefont
  {Leemans}}]{gonsalves2011}%
  \BibitemOpen
  \bibfield  {author} {\bibinfo {author} {\bibfnamefont {A.~J.}\ \bibnamefont
  {Gonsalves}}, \bibinfo {author} {\bibfnamefont {K.}~\bibnamefont {Nakamura}},
  \bibinfo {author} {\bibfnamefont {C.}~\bibnamefont {Lin}}, \bibinfo {author}
  {\bibfnamefont {D.}~\bibnamefont {Panasenko}}, \bibinfo {author}
  {\bibfnamefont {S.}~\bibnamefont {Shiraishi}}, \bibinfo {author}
  {\bibfnamefont {T.}~\bibnamefont {Sokollik}}, \bibinfo {author}
  {\bibfnamefont {C.}~\bibnamefont {Benedetti}}, \bibinfo {author}
  {\bibfnamefont {C.~B.}\ \bibnamefont {Schroeder}}, \bibinfo {author}
  {\bibfnamefont {C.~G.~R.}\ \bibnamefont {Geddes}}, \bibinfo {author}
  {\bibfnamefont {J.}~\bibnamefont {{van Tilborg}}}, \bibinfo {author}
  {\bibfnamefont {J.}~\bibnamefont {Osterhoff}}, \bibinfo {author}
  {\bibfnamefont {E.}~\bibnamefont {Esarey}}, \bibinfo {author} {\bibfnamefont
  {C.}~\bibnamefont {Toth}},\ and\ \bibinfo {author} {\bibfnamefont {W.~P.}\
  \bibnamefont {Leemans}},\ }\bibfield  {title} {\bibinfo {title} {Tunable
  laser plasma accelerator based on longitudinal density tailoring},\ }\href
  {https://doi.org/10.1038/NPHYS2071} {\bibfield  {journal} {\bibinfo
  {journal} {Nat. Phys.}\ }\textbf {\bibinfo {volume} {7}},\ \bibinfo {pages}
  {862} (\bibinfo {year} {2011})}\BibitemShut {NoStop}%
\bibitem [{\citenamefont {Vieira}\ \emph
  {et~al.}(2011{\natexlab{a}})\citenamefont {Vieira}, \citenamefont {Martins},
  \citenamefont {Pathak}, \citenamefont {Fonseca}, \citenamefont {Mori},\ and\
  \citenamefont {Silva}}]{vieira2011magnetic}%
  \BibitemOpen
  \bibfield  {author} {\bibinfo {author} {\bibfnamefont {J.}~\bibnamefont
  {Vieira}}, \bibinfo {author} {\bibfnamefont {S.~F.}\ \bibnamefont {Martins}},
  \bibinfo {author} {\bibfnamefont {V.~B.}\ \bibnamefont {Pathak}}, \bibinfo
  {author} {\bibfnamefont {R.~A.}\ \bibnamefont {Fonseca}}, \bibinfo {author}
  {\bibfnamefont {W.~B.}\ \bibnamefont {Mori}},\ and\ \bibinfo {author}
  {\bibfnamefont {L.~O.}\ \bibnamefont {Silva}},\ }\bibfield  {title} {\bibinfo
  {title} {Magnetic control of particle injection in plasma based
  accelerators},\ }\href {https://doi.org/10.1103/PhysRevLett.106.225001}
  {\bibfield  {journal} {\bibinfo  {journal} {Phys. Rev. Lett.}\ }\textbf
  {\bibinfo {volume} {106}},\ \bibinfo {pages} {225001} (\bibinfo {year}
  {2011}{\natexlab{a}})}\BibitemShut {NoStop}%
\bibitem [{\citenamefont {Zhao}\ \emph {et~al.}(2018)\citenamefont {Zhao},
  \citenamefont {Weng}, \citenamefont {Sheng}, \citenamefont {Chen},
  \citenamefont {Zhang}, \citenamefont {Mori}, \citenamefont {Hidding},
  \citenamefont {Jaroszynski},\ and\ \citenamefont {Zhang}}]{zhao2018}%
  \BibitemOpen
  \bibfield  {author} {\bibinfo {author} {\bibfnamefont {Q.}~\bibnamefont
  {Zhao}}, \bibinfo {author} {\bibfnamefont {S.~M.}\ \bibnamefont {Weng}},
  \bibinfo {author} {\bibfnamefont {Z.~M.}\ \bibnamefont {Sheng}}, \bibinfo
  {author} {\bibfnamefont {M.}~\bibnamefont {Chen}}, \bibinfo {author}
  {\bibfnamefont {G.~B.}\ \bibnamefont {Zhang}}, \bibinfo {author}
  {\bibfnamefont {W.~B.}\ \bibnamefont {Mori}}, \bibinfo {author}
  {\bibfnamefont {B.}~\bibnamefont {Hidding}}, \bibinfo {author} {\bibfnamefont
  {D.~A.}\ \bibnamefont {Jaroszynski}},\ and\ \bibinfo {author} {\bibfnamefont
  {J.}~\bibnamefont {Zhang}},\ }\bibfield  {title} {\bibinfo {title}
  {Ionization injection in a laser wakefield accelerator subject to a
  transverse magnetic field},\ }\href
  {https://doi.org/10.1088/1367-2630/aac926} {\bibfield  {journal} {\bibinfo
  {journal} {New J. Phys.}\ }\textbf {\bibinfo {volume} {20}},\ \bibinfo
  {pages} {063031} (\bibinfo {year} {2018})}\BibitemShut {NoStop}%
\bibitem [{\citenamefont {Luttikhof}\ \emph {et~al.}(2010)\citenamefont
  {Luttikhof}, \citenamefont {Khachatryan}, \citenamefont {{van Goor}},\ and\
  \citenamefont {Boller}}]{luttikhof2010}%
  \BibitemOpen
  \bibfield  {author} {\bibinfo {author} {\bibfnamefont {M.~J.~H.}\
  \bibnamefont {Luttikhof}}, \bibinfo {author} {\bibfnamefont {A.~G.}\
  \bibnamefont {Khachatryan}}, \bibinfo {author} {\bibfnamefont {F.~A.}\
  \bibnamefont {{van Goor}}},\ and\ \bibinfo {author} {\bibfnamefont {K.-J.}\
  \bibnamefont {Boller}},\ }\bibfield  {title} {\bibinfo {title} {Generating
  {{Ultrarelativistic Attosecond Electron Bunches}} with {{Laser Wakefield
  Accelerators}}},\ }\href {https://doi.org/10.1103/PhysRevLett.105.124801}
  {\bibfield  {journal} {\bibinfo  {journal} {Phys. Rev. Lett.}\ }\textbf
  {\bibinfo {volume} {105}},\ \bibinfo {pages} {124801} (\bibinfo {year}
  {2010})}\BibitemShut {NoStop}%
\bibitem [{\citenamefont {Tooley}\ \emph {et~al.}(2017)\citenamefont {Tooley},
  \citenamefont {Ersfeld}, \citenamefont {Yoffe}, \citenamefont {Noble},
  \citenamefont {Brunetti}, \citenamefont {Sheng}, \citenamefont {Islam},\ and\
  \citenamefont {Jaroszynski}}]{tooley2017}%
  \BibitemOpen
  \bibfield  {author} {\bibinfo {author} {\bibfnamefont {M.~P.}\ \bibnamefont
  {Tooley}}, \bibinfo {author} {\bibfnamefont {B.}~\bibnamefont {Ersfeld}},
  \bibinfo {author} {\bibfnamefont {S.~R.}\ \bibnamefont {Yoffe}}, \bibinfo
  {author} {\bibfnamefont {A.}~\bibnamefont {Noble}}, \bibinfo {author}
  {\bibfnamefont {E.}~\bibnamefont {Brunetti}}, \bibinfo {author}
  {\bibfnamefont {Z.~M.}\ \bibnamefont {Sheng}}, \bibinfo {author}
  {\bibfnamefont {M.~R.}\ \bibnamefont {Islam}},\ and\ \bibinfo {author}
  {\bibfnamefont {D.~A.}\ \bibnamefont {Jaroszynski}},\ }\bibfield  {title}
  {\bibinfo {title} {Towards attosecond high-energy electron bunches:
  Controlling self-injection in laser-wakefield accelerators through
  plasma-density modulation},\ }\href
  {https://doi.org/10.1103/PhysRevLett.119.044801} {\bibfield  {journal}
  {\bibinfo  {journal} {Phys. Rev. Lett.}\ }\textbf {\bibinfo {volume} {119}},\
  \bibinfo {pages} {044801} (\bibinfo {year} {2017})}\BibitemShut {NoStop}%
\bibitem [{\citenamefont {Zhao}\ \emph {et~al.}(2019)\citenamefont {Zhao},
  \citenamefont {Weng}, \citenamefont {Chen}, \citenamefont {Zeng},
  \citenamefont {Hidding}, \citenamefont {Jaroszynski}, \citenamefont
  {Assmann},\ and\ \citenamefont {Sheng}}]{zhao2019}%
  \BibitemOpen
  \bibfield  {author} {\bibinfo {author} {\bibfnamefont {Q.}~\bibnamefont
  {Zhao}}, \bibinfo {author} {\bibfnamefont {S.}~\bibnamefont {Weng}}, \bibinfo
  {author} {\bibfnamefont {M.}~\bibnamefont {Chen}}, \bibinfo {author}
  {\bibfnamefont {M.}~\bibnamefont {Zeng}}, \bibinfo {author} {\bibfnamefont
  {B.}~\bibnamefont {Hidding}}, \bibinfo {author} {\bibfnamefont
  {D.}~\bibnamefont {Jaroszynski}}, \bibinfo {author} {\bibfnamefont
  {R.}~\bibnamefont {Assmann}},\ and\ \bibinfo {author} {\bibfnamefont
  {Z.}~\bibnamefont {Sheng}},\ }\bibfield  {title} {\bibinfo {title}
  {Sub-femtosecond electron bunches in laser wakefield acceleration via
  injection suppression with a magnetic field},\ }\href@noop {} {\bibfield
  {journal} {\bibinfo  {journal} {Plasma Phys. Control. Fusion}\ }\textbf
  {\bibinfo {volume} {61}},\ \bibinfo {pages} {085015} (\bibinfo {year}
  {2019})}\BibitemShut {NoStop}%
\bibitem [{\citenamefont {Kim}\ \emph {et~al.}(2021)\citenamefont {Kim},
  \citenamefont {Wang}, \citenamefont {Khudik},\ and\ \citenamefont
  {Shvets}}]{kim2021}%
  \BibitemOpen
  \bibfield  {author} {\bibinfo {author} {\bibfnamefont {J.}~\bibnamefont
  {Kim}}, \bibinfo {author} {\bibfnamefont {T.}~\bibnamefont {Wang}}, \bibinfo
  {author} {\bibfnamefont {V.}~\bibnamefont {Khudik}},\ and\ \bibinfo {author}
  {\bibfnamefont {G.}~\bibnamefont {Shvets}},\ }\bibfield  {title} {\bibinfo
  {title} {Subfemtosecond {{Wakefield Injector}} and {{Accelerator Based}} on
  an {{Undulating Plasma Bubble Controlled}} by a {{Laser Phase}}},\ }\href
  {https://doi.org/10.1103/PhysRevLett.127.164801} {\bibfield  {journal}
  {\bibinfo  {journal} {Phys. Rev. Lett.}\ }\textbf {\bibinfo {volume} {127}},\
  \bibinfo {pages} {164801} (\bibinfo {year} {2021})}\BibitemShut {NoStop}%
\bibitem [{\citenamefont {Baum}\ and\ \citenamefont {Zewail}(2007)}]{baum2007}%
  \BibitemOpen
  \bibfield  {author} {\bibinfo {author} {\bibfnamefont {P.}~\bibnamefont
  {Baum}}\ and\ \bibinfo {author} {\bibfnamefont {A.~H.}\ \bibnamefont
  {Zewail}},\ }\bibfield  {title} {\bibinfo {title} {Attosecond electron pulses
  for {{4D}} diffraction and microscopy},\ }\href
  {https://doi.org/10.1073/pnas.0709019104} {\bibfield  {journal} {\bibinfo
  {journal} {Proc. Natl. Acad. Sci. U.S.A.}\ }\textbf {\bibinfo {volume}
  {104}},\ \bibinfo {pages} {18409} (\bibinfo {year} {2007})}\BibitemShut
  {NoStop}%
\bibitem [{\citenamefont {Krausz}\ and\ \citenamefont
  {Ivanov}(2009)}]{krausz2009}%
  \BibitemOpen
  \bibfield  {author} {\bibinfo {author} {\bibfnamefont {F.}~\bibnamefont
  {Krausz}}\ and\ \bibinfo {author} {\bibfnamefont {M.}~\bibnamefont
  {Ivanov}},\ }\bibfield  {title} {\bibinfo {title} {Attosecond physics},\
  }\href {https://doi.org/10.1103/RevModPhys.81.163} {\bibfield  {journal}
  {\bibinfo  {journal} {Rev. Mod. Phys.}\ }\textbf {\bibinfo {volume} {81}},\
  \bibinfo {pages} {163} (\bibinfo {year} {2009})}\BibitemShut {NoStop}%
\bibitem [{\citenamefont {Corde}\ \emph {et~al.}(2013)\citenamefont {Corde},
  \citenamefont {Ta~Phuoc}, \citenamefont {Lambert}, \citenamefont {Fitour},
  \citenamefont {Malka}, \citenamefont {Rousse}, \citenamefont {Beck},\ and\
  \citenamefont {Lefebvre}}]{corde2013}%
  \BibitemOpen
  \bibfield  {author} {\bibinfo {author} {\bibfnamefont {S.}~\bibnamefont
  {Corde}}, \bibinfo {author} {\bibfnamefont {K.}~\bibnamefont {Ta~Phuoc}},
  \bibinfo {author} {\bibfnamefont {G.}~\bibnamefont {Lambert}}, \bibinfo
  {author} {\bibfnamefont {R.}~\bibnamefont {Fitour}}, \bibinfo {author}
  {\bibfnamefont {V.}~\bibnamefont {Malka}}, \bibinfo {author} {\bibfnamefont
  {A.}~\bibnamefont {Rousse}}, \bibinfo {author} {\bibfnamefont
  {A.}~\bibnamefont {Beck}},\ and\ \bibinfo {author} {\bibfnamefont
  {E.}~\bibnamefont {Lefebvre}},\ }\bibfield  {title} {\bibinfo {title}
  {Femtosecond x rays from laser-plasma accelerators},\ }\href
  {https://doi.org/10.1103/RevModPhys.85.1} {\bibfield  {journal} {\bibinfo
  {journal} {Rev. Mod. Phys.}\ }\textbf {\bibinfo {volume} {85}},\ \bibinfo
  {pages} {1} (\bibinfo {year} {2013})}\BibitemShut {NoStop}%
\bibitem [{\citenamefont {Miller}(2014)}]{miller2014}%
  \BibitemOpen
  \bibfield  {author} {\bibinfo {author} {\bibfnamefont {R.~J.~D.}\
  \bibnamefont {Miller}},\ }\bibfield  {title} {\bibinfo {title} {Femtosecond
  {{Crystallography}} with {{Ultrabright Electrons}} and {{X-rays}}:
  {{Capturing Chemistry}} in {{Action}}},\ }\href
  {https://doi.org/10.1126/science.1248488} {\bibfield  {journal} {\bibinfo
  {journal} {Science}\ }\textbf {\bibinfo {volume} {343}},\ \bibinfo {pages}
  {1108} (\bibinfo {year} {2014})}\BibitemShut {NoStop}%
\bibitem [{\citenamefont {Morimoto}\ and\ \citenamefont
  {Baum}(2018)}]{morimoto2018}%
  \BibitemOpen
  \bibfield  {author} {\bibinfo {author} {\bibfnamefont {Y.}~\bibnamefont
  {Morimoto}}\ and\ \bibinfo {author} {\bibfnamefont {P.}~\bibnamefont
  {Baum}},\ }\bibfield  {title} {\bibinfo {title} {Diffraction and microscopy
  with attosecond electron pulse trains},\ }\href
  {https://doi.org/10.1038/s41567-017-0007-6} {\bibfield  {journal} {\bibinfo
  {journal} {Nature Phys.}\ }\textbf {\bibinfo {volume} {14}},\ \bibinfo
  {pages} {252} (\bibinfo {year} {2018})}\BibitemShut {NoStop}%
\bibitem [{\citenamefont {Luo}\ \emph {et~al.}(2018)\citenamefont {Luo},
  \citenamefont {Chen}, \citenamefont {Wu}, \citenamefont {Weng}, \citenamefont
  {Sheng}, \citenamefont {Schroeder}, \citenamefont {Jaroszynski},
  \citenamefont {Esarey}, \citenamefont {Leemans}, \citenamefont {Mori},\ and\
  \citenamefont {Zhang}}]{luo2018}%
  \BibitemOpen
  \bibfield  {author} {\bibinfo {author} {\bibfnamefont {J.}~\bibnamefont
  {Luo}}, \bibinfo {author} {\bibfnamefont {M.}~\bibnamefont {Chen}}, \bibinfo
  {author} {\bibfnamefont {W.~Y.}\ \bibnamefont {Wu}}, \bibinfo {author}
  {\bibfnamefont {S.~M.}\ \bibnamefont {Weng}}, \bibinfo {author}
  {\bibfnamefont {Z.~M.}\ \bibnamefont {Sheng}}, \bibinfo {author}
  {\bibfnamefont {C.~B.}\ \bibnamefont {Schroeder}}, \bibinfo {author}
  {\bibfnamefont {D.~A.}\ \bibnamefont {Jaroszynski}}, \bibinfo {author}
  {\bibfnamefont {E.}~\bibnamefont {Esarey}}, \bibinfo {author} {\bibfnamefont
  {W.~P.}\ \bibnamefont {Leemans}}, \bibinfo {author} {\bibfnamefont {W.~B.}\
  \bibnamefont {Mori}},\ and\ \bibinfo {author} {\bibfnamefont
  {J.}~\bibnamefont {Zhang}},\ }\bibfield  {title} {\bibinfo {title}
  {Multistage coupling of laser-wakefield accelerators with curved plasma
  channels},\ }\href {https://doi.org/10.1103/PhysRevLett.120.154801}
  {\bibfield  {journal} {\bibinfo  {journal} {Phys. Rev. Lett.}\ }\textbf
  {\bibinfo {volume} {120}},\ \bibinfo {pages} {154801} (\bibinfo {year}
  {2018})}\BibitemShut {NoStop}%
\bibitem [{\citenamefont {Zhu}\ \emph {et~al.}(2023)\citenamefont {Zhu},
  \citenamefont {Li}, \citenamefont {Liu}, \citenamefont {Li}, \citenamefont
  {Bi}, \citenamefont {Ge}, \citenamefont {Deng}, \citenamefont {Zhang},
  \citenamefont {Cui}, \citenamefont {Lu}, \citenamefont {Yan}, \citenamefont
  {Yuan}, \citenamefont {Chen}, \citenamefont {Cao}, \citenamefont {Liu},
  \citenamefont {Sheng}, \citenamefont {Chen},\ and\ \citenamefont
  {Zhang}}]{zhu2023}%
  \BibitemOpen
  \bibfield  {author} {\bibinfo {author} {\bibfnamefont {X.}~\bibnamefont
  {Zhu}}, \bibinfo {author} {\bibfnamefont {B.}~\bibnamefont {Li}}, \bibinfo
  {author} {\bibfnamefont {F.}~\bibnamefont {Liu}}, \bibinfo {author}
  {\bibfnamefont {J.}~\bibnamefont {Li}}, \bibinfo {author} {\bibfnamefont
  {Z.}~\bibnamefont {Bi}}, \bibinfo {author} {\bibfnamefont {X.}~\bibnamefont
  {Ge}}, \bibinfo {author} {\bibfnamefont {H.}~\bibnamefont {Deng}}, \bibinfo
  {author} {\bibfnamefont {Z.}~\bibnamefont {Zhang}}, \bibinfo {author}
  {\bibfnamefont {P.}~\bibnamefont {Cui}}, \bibinfo {author} {\bibfnamefont
  {L.}~\bibnamefont {Lu}}, \bibinfo {author} {\bibfnamefont {W.}~\bibnamefont
  {Yan}}, \bibinfo {author} {\bibfnamefont {X.}~\bibnamefont {Yuan}}, \bibinfo
  {author} {\bibfnamefont {L.}~\bibnamefont {Chen}}, \bibinfo {author}
  {\bibfnamefont {Q.}~\bibnamefont {Cao}}, \bibinfo {author} {\bibfnamefont
  {Z.}~\bibnamefont {Liu}}, \bibinfo {author} {\bibfnamefont {Z.}~\bibnamefont
  {Sheng}}, \bibinfo {author} {\bibfnamefont {M.}~\bibnamefont {Chen}},\ and\
  \bibinfo {author} {\bibfnamefont {J.}~\bibnamefont {Zhang}},\ }\bibfield
  {title} {\bibinfo {title} {Experimental demonstration of laser guiding and
  wakefield acceleration in a curved plasma channel},\ }\href
  {https://doi.org/10.1103/PhysRevLett.130.215001} {\bibfield  {journal}
  {\bibinfo  {journal} {Phys. Rev. Lett.}\ }\textbf {\bibinfo {volume} {130}},\
  \bibinfo {pages} {215001} (\bibinfo {year} {2023})}\BibitemShut {NoStop}%
\bibitem [{\citenamefont {Di~Mitri}\ and\ \citenamefont
  {Cornacchia}(2014)}]{dimitri2014}%
  \BibitemOpen
  \bibfield  {author} {\bibinfo {author} {\bibfnamefont {S.}~\bibnamefont
  {Di~Mitri}}\ and\ \bibinfo {author} {\bibfnamefont {M.}~\bibnamefont
  {Cornacchia}},\ }\bibfield  {title} {\bibinfo {title} {Electron beam
  brightness in linac drivers for free-electron-lasers},\ }\href
  {https://doi.org/10.1016/j.physrep.2014.01.005} {\bibfield  {journal}
  {\bibinfo  {journal} {Phys. Rep.}\ }\textbf {\bibinfo {volume} {539}},\
  \bibinfo {pages} {1} (\bibinfo {year} {2014})}\BibitemShut {NoStop}%
\bibitem [{\citenamefont {Wang}\ \emph {et~al.}(2016)\citenamefont {Wang},
  \citenamefont {Li}, \citenamefont {Liu}, \citenamefont {Zhang}, \citenamefont
  {Qi}, \citenamefont {Yu}, \citenamefont {Liu}, \citenamefont {Fang},
  \citenamefont {Qin}, \citenamefont {Wang}, \citenamefont {Xu}, \citenamefont
  {Wu}, \citenamefont {Leng}, \citenamefont {Li},\ and\ \citenamefont
  {Xu}}]{wang2016}%
  \BibitemOpen
  \bibfield  {author} {\bibinfo {author} {\bibfnamefont {W.~T.}\ \bibnamefont
  {Wang}}, \bibinfo {author} {\bibfnamefont {W.~T.}\ \bibnamefont {Li}},
  \bibinfo {author} {\bibfnamefont {J.~S.}\ \bibnamefont {Liu}}, \bibinfo
  {author} {\bibfnamefont {Z.~J.}\ \bibnamefont {Zhang}}, \bibinfo {author}
  {\bibfnamefont {R.}~\bibnamefont {Qi}}, \bibinfo {author} {\bibfnamefont
  {C.~H.}\ \bibnamefont {Yu}}, \bibinfo {author} {\bibfnamefont {J.~Q.}\
  \bibnamefont {Liu}}, \bibinfo {author} {\bibfnamefont {M.}~\bibnamefont
  {Fang}}, \bibinfo {author} {\bibfnamefont {Z.~Y.}\ \bibnamefont {Qin}},
  \bibinfo {author} {\bibfnamefont {C.}~\bibnamefont {Wang}}, \bibinfo {author}
  {\bibfnamefont {Y.}~\bibnamefont {Xu}}, \bibinfo {author} {\bibfnamefont
  {F.~X.}\ \bibnamefont {Wu}}, \bibinfo {author} {\bibfnamefont {Y.~X.}\
  \bibnamefont {Leng}}, \bibinfo {author} {\bibfnamefont {R.~X.}\ \bibnamefont
  {Li}},\ and\ \bibinfo {author} {\bibfnamefont {Z.~Z.}\ \bibnamefont {Xu}},\
  }\bibfield  {title} {\bibinfo {title} {High-{{Brightness High-Energy Electron
  Beams}} from a {{Laser Wakefield Accelerator}} via {{Energy Chirp
  Control}}},\ }\href {https://doi.org/10.1103/PhysRevLett.117.124801}
  {\bibfield  {journal} {\bibinfo  {journal} {Phys. Rev. Lett.}\ }\textbf
  {\bibinfo {volume} {117}},\ \bibinfo {pages} {124801} (\bibinfo {year}
  {2016})}\BibitemShut {NoStop}%
\bibitem [{\citenamefont {Huang}\ \emph {et~al.}(2012)\citenamefont {Huang},
  \citenamefont {Ding},\ and\ \citenamefont {Schroeder}}]{huang2012}%
  \BibitemOpen
  \bibfield  {author} {\bibinfo {author} {\bibfnamefont {Z.}~\bibnamefont
  {Huang}}, \bibinfo {author} {\bibfnamefont {Y.}~\bibnamefont {Ding}},\ and\
  \bibinfo {author} {\bibfnamefont {C.~B.}\ \bibnamefont {Schroeder}},\
  }\bibfield  {title} {\bibinfo {title} {Compact {{X-ray Free-Electron Laser}}
  from a {{Laser-Plasma Accelerator Using}} a {{Transverse-Gradient
  Undulator}}},\ }\href {https://doi.org/10.1103/PhysRevLett.109.204801}
  {\bibfield  {journal} {\bibinfo  {journal} {Phys. Rev. Lett.}\ }\textbf
  {\bibinfo {volume} {109}},\ \bibinfo {pages} {204801} (\bibinfo {year}
  {2012})}\BibitemShut {NoStop}%
\bibitem [{\citenamefont {Wang}\ \emph {et~al.}(2021)\citenamefont {Wang},
  \citenamefont {Feng}, \citenamefont {Ke}, \citenamefont {Yu}, \citenamefont
  {Xu}, \citenamefont {Qi}, \citenamefont {Chen}, \citenamefont {Qin},
  \citenamefont {Zhang}, \citenamefont {Fang}, \citenamefont {Liu},
  \citenamefont {Jiang}, \citenamefont {Wang}, \citenamefont {Wang},
  \citenamefont {Yang}, \citenamefont {Wu}, \citenamefont {Leng}, \citenamefont
  {Liu}, \citenamefont {Li},\ and\ \citenamefont {Xu}}]{wang2021}%
  \BibitemOpen
  \bibfield  {author} {\bibinfo {author} {\bibfnamefont {W.}~\bibnamefont
  {Wang}}, \bibinfo {author} {\bibfnamefont {K.}~\bibnamefont {Feng}}, \bibinfo
  {author} {\bibfnamefont {L.}~\bibnamefont {Ke}}, \bibinfo {author}
  {\bibfnamefont {C.}~\bibnamefont {Yu}}, \bibinfo {author} {\bibfnamefont
  {Y.}~\bibnamefont {Xu}}, \bibinfo {author} {\bibfnamefont {R.}~\bibnamefont
  {Qi}}, \bibinfo {author} {\bibfnamefont {Y.}~\bibnamefont {Chen}}, \bibinfo
  {author} {\bibfnamefont {Z.}~\bibnamefont {Qin}}, \bibinfo {author}
  {\bibfnamefont {Z.}~\bibnamefont {Zhang}}, \bibinfo {author} {\bibfnamefont
  {M.}~\bibnamefont {Fang}}, \bibinfo {author} {\bibfnamefont {J.}~\bibnamefont
  {Liu}}, \bibinfo {author} {\bibfnamefont {K.}~\bibnamefont {Jiang}}, \bibinfo
  {author} {\bibfnamefont {H.}~\bibnamefont {Wang}}, \bibinfo {author}
  {\bibfnamefont {C.}~\bibnamefont {Wang}}, \bibinfo {author} {\bibfnamefont
  {X.}~\bibnamefont {Yang}}, \bibinfo {author} {\bibfnamefont {F.}~\bibnamefont
  {Wu}}, \bibinfo {author} {\bibfnamefont {Y.}~\bibnamefont {Leng}}, \bibinfo
  {author} {\bibfnamefont {J.}~\bibnamefont {Liu}}, \bibinfo {author}
  {\bibfnamefont {R.}~\bibnamefont {Li}},\ and\ \bibinfo {author}
  {\bibfnamefont {Z.}~\bibnamefont {Xu}},\ }\bibfield  {title} {\bibinfo
  {title} {Free-electron lasing at 27 nanometres based on a laser wakefield
  accelerator},\ }\href {https://doi.org/10.1038/s41586-021-03678-x} {\bibfield
   {journal} {\bibinfo  {journal} {Nature}\ }\textbf {\bibinfo {volume}
  {595}},\ \bibinfo {pages} {516} (\bibinfo {year} {2021})}\BibitemShut
  {NoStop}%
\bibitem [{\citenamefont {Pompili}\ \emph {et~al.}(2022)\citenamefont
  {Pompili}, \citenamefont {Alesini}, \citenamefont {Anania}, \citenamefont
  {Arjmand}, \citenamefont {Behtouei}, \citenamefont {Bellaveglia},
  \citenamefont {Biagioni}, \citenamefont {Buonomo}, \citenamefont {Cardelli},
  \citenamefont {Carpanese}, \citenamefont {Chiadroni}, \citenamefont
  {Cianchi}, \citenamefont {Costa}, \citenamefont {Del~Dotto}, \citenamefont
  {Del~Giorno}, \citenamefont {Dipace}, \citenamefont {Doria}, \citenamefont
  {Filippi}, \citenamefont {Galletti}, \citenamefont {Giannessi}, \citenamefont
  {Giribono}, \citenamefont {Iovine}, \citenamefont {Lollo}, \citenamefont
  {Mostacci}, \citenamefont {Nguyen}, \citenamefont {Opromolla}, \citenamefont
  {Di~Palma}, \citenamefont {Pellegrino}, \citenamefont {Petralia},
  \citenamefont {Petrillo}, \citenamefont {Piersanti}, \citenamefont
  {Di~Pirro}, \citenamefont {Romeo}, \citenamefont {Rossi}, \citenamefont
  {Scifo}, \citenamefont {Selce}, \citenamefont {Shpakov}, \citenamefont
  {Stella}, \citenamefont {Vaccarezza}, \citenamefont {Villa}, \citenamefont
  {Zigler},\ and\ \citenamefont {Ferrario}}]{pompili2022}%
  \BibitemOpen
  \bibfield  {author} {\bibinfo {author} {\bibfnamefont {R.}~\bibnamefont
  {Pompili}}, \bibinfo {author} {\bibfnamefont {D.}~\bibnamefont {Alesini}},
  \bibinfo {author} {\bibfnamefont {M.~P.}\ \bibnamefont {Anania}}, \bibinfo
  {author} {\bibfnamefont {S.}~\bibnamefont {Arjmand}}, \bibinfo {author}
  {\bibfnamefont {M.}~\bibnamefont {Behtouei}}, \bibinfo {author}
  {\bibfnamefont {M.}~\bibnamefont {Bellaveglia}}, \bibinfo {author}
  {\bibfnamefont {A.}~\bibnamefont {Biagioni}}, \bibinfo {author}
  {\bibfnamefont {B.}~\bibnamefont {Buonomo}}, \bibinfo {author} {\bibfnamefont
  {F.}~\bibnamefont {Cardelli}}, \bibinfo {author} {\bibfnamefont
  {M.}~\bibnamefont {Carpanese}}, \bibinfo {author} {\bibfnamefont
  {E.}~\bibnamefont {Chiadroni}}, \bibinfo {author} {\bibfnamefont
  {A.}~\bibnamefont {Cianchi}}, \bibinfo {author} {\bibfnamefont
  {G.}~\bibnamefont {Costa}}, \bibinfo {author} {\bibfnamefont
  {A.}~\bibnamefont {Del~Dotto}}, \bibinfo {author} {\bibfnamefont
  {M.}~\bibnamefont {Del~Giorno}}, \bibinfo {author} {\bibfnamefont
  {F.}~\bibnamefont {Dipace}}, \bibinfo {author} {\bibfnamefont
  {A.}~\bibnamefont {Doria}}, \bibinfo {author} {\bibfnamefont
  {F.}~\bibnamefont {Filippi}}, \bibinfo {author} {\bibfnamefont
  {M.}~\bibnamefont {Galletti}}, \bibinfo {author} {\bibfnamefont
  {L.}~\bibnamefont {Giannessi}}, \bibinfo {author} {\bibfnamefont
  {A.}~\bibnamefont {Giribono}}, \bibinfo {author} {\bibfnamefont
  {P.}~\bibnamefont {Iovine}}, \bibinfo {author} {\bibfnamefont
  {V.}~\bibnamefont {Lollo}}, \bibinfo {author} {\bibfnamefont
  {A.}~\bibnamefont {Mostacci}}, \bibinfo {author} {\bibfnamefont
  {F.}~\bibnamefont {Nguyen}}, \bibinfo {author} {\bibfnamefont
  {M.}~\bibnamefont {Opromolla}}, \bibinfo {author} {\bibfnamefont
  {E.}~\bibnamefont {Di~Palma}}, \bibinfo {author} {\bibfnamefont
  {L.}~\bibnamefont {Pellegrino}}, \bibinfo {author} {\bibfnamefont
  {A.}~\bibnamefont {Petralia}}, \bibinfo {author} {\bibfnamefont
  {V.}~\bibnamefont {Petrillo}}, \bibinfo {author} {\bibfnamefont
  {L.}~\bibnamefont {Piersanti}}, \bibinfo {author} {\bibfnamefont
  {G.}~\bibnamefont {Di~Pirro}}, \bibinfo {author} {\bibfnamefont
  {S.}~\bibnamefont {Romeo}}, \bibinfo {author} {\bibfnamefont {A.~R.}\
  \bibnamefont {Rossi}}, \bibinfo {author} {\bibfnamefont {J.}~\bibnamefont
  {Scifo}}, \bibinfo {author} {\bibfnamefont {A.}~\bibnamefont {Selce}},
  \bibinfo {author} {\bibfnamefont {V.}~\bibnamefont {Shpakov}}, \bibinfo
  {author} {\bibfnamefont {A.}~\bibnamefont {Stella}}, \bibinfo {author}
  {\bibfnamefont {C.}~\bibnamefont {Vaccarezza}}, \bibinfo {author}
  {\bibfnamefont {F.}~\bibnamefont {Villa}}, \bibinfo {author} {\bibfnamefont
  {A.}~\bibnamefont {Zigler}},\ and\ \bibinfo {author} {\bibfnamefont
  {M.}~\bibnamefont {Ferrario}},\ }\bibfield  {title} {\bibinfo {title}
  {Free-electron lasing with compact beam-driven plasma wakefield
  accelerator},\ }\href {https://doi.org/10.1038/s41586-022-04589-1} {\bibfield
   {journal} {\bibinfo  {journal} {Nature}\ }\textbf {\bibinfo {volume}
  {605}},\ \bibinfo {pages} {659} (\bibinfo {year} {2022})}\BibitemShut
  {NoStop}%
\bibitem [{\citenamefont {Labat}\ \emph {et~al.}(2023)\citenamefont {Labat},
  \citenamefont {Cabadağ}, \citenamefont {Ghaith}, \citenamefont {Irman},
  \citenamefont {Berlioux}, \citenamefont {Berteaud}, \citenamefont {Blache},
  \citenamefont {Bock}, \citenamefont {Bouvet}, \citenamefont {Briquez},
  \citenamefont {Chang}, \citenamefont {Corde}, \citenamefont {Debus},
  \citenamefont {De~Oliveira}, \citenamefont {Duval}, \citenamefont {Dietrich},
  \citenamefont {El~Ajjouri}, \citenamefont {Eisenmann}, \citenamefont
  {Gautier}, \citenamefont {Gebhardt}, \citenamefont {Grams}, \citenamefont
  {Helbig}, \citenamefont {Herbeaux}, \citenamefont {Hubert}, \citenamefont
  {Kitegi}, \citenamefont {Kononenko}, \citenamefont {Kuntzsch}, \citenamefont
  {LaBerge}, \citenamefont {Lê}, \citenamefont {Leluan}, \citenamefont
  {Loulergue}, \citenamefont {Malka}, \citenamefont {Marteau}, \citenamefont
  {Guyen}, \citenamefont {Oumbarek-Espinos}, \citenamefont {Pausch},
  \citenamefont {Pereira}, \citenamefont {Püschel}, \citenamefont {Ricaud},
  \citenamefont {Rommeluere}, \citenamefont {Roussel}, \citenamefont
  {Rousseau}, \citenamefont {Schöbel}, \citenamefont {Sebdaoui}, \citenamefont
  {Steiniger}, \citenamefont {Tavakoli}, \citenamefont {Thaury}, \citenamefont
  {Ufer}, \citenamefont {Valléau}, \citenamefont {Vandenberghe}, \citenamefont
  {Vétéran}, \citenamefont {Schramm},\ and\ \citenamefont
  {Couprie}}]{labat2023}%
  \BibitemOpen
  \bibfield  {author} {\bibinfo {author} {\bibfnamefont {M.}~\bibnamefont
  {Labat}}, \bibinfo {author} {\bibfnamefont {J.~C.}\ \bibnamefont {Cabadağ}},
  \bibinfo {author} {\bibfnamefont {A.}~\bibnamefont {Ghaith}}, \bibinfo
  {author} {\bibfnamefont {A.}~\bibnamefont {Irman}}, \bibinfo {author}
  {\bibfnamefont {A.}~\bibnamefont {Berlioux}}, \bibinfo {author}
  {\bibfnamefont {P.}~\bibnamefont {Berteaud}}, \bibinfo {author}
  {\bibfnamefont {F.}~\bibnamefont {Blache}}, \bibinfo {author} {\bibfnamefont
  {S.}~\bibnamefont {Bock}}, \bibinfo {author} {\bibfnamefont {F.}~\bibnamefont
  {Bouvet}}, \bibinfo {author} {\bibfnamefont {F.}~\bibnamefont {Briquez}},
  \bibinfo {author} {\bibfnamefont {Y.-Y.}\ \bibnamefont {Chang}}, \bibinfo
  {author} {\bibfnamefont {S.}~\bibnamefont {Corde}}, \bibinfo {author}
  {\bibfnamefont {A.}~\bibnamefont {Debus}}, \bibinfo {author} {\bibfnamefont
  {C.}~\bibnamefont {De~Oliveira}}, \bibinfo {author} {\bibfnamefont {J.-P.}\
  \bibnamefont {Duval}}, \bibinfo {author} {\bibfnamefont {Y.}~\bibnamefont
  {Dietrich}}, \bibinfo {author} {\bibfnamefont {M.}~\bibnamefont
  {El~Ajjouri}}, \bibinfo {author} {\bibfnamefont {C.}~\bibnamefont
  {Eisenmann}}, \bibinfo {author} {\bibfnamefont {J.}~\bibnamefont {Gautier}},
  \bibinfo {author} {\bibfnamefont {R.}~\bibnamefont {Gebhardt}}, \bibinfo
  {author} {\bibfnamefont {S.}~\bibnamefont {Grams}}, \bibinfo {author}
  {\bibfnamefont {U.}~\bibnamefont {Helbig}}, \bibinfo {author} {\bibfnamefont
  {C.}~\bibnamefont {Herbeaux}}, \bibinfo {author} {\bibfnamefont
  {N.}~\bibnamefont {Hubert}}, \bibinfo {author} {\bibfnamefont
  {C.}~\bibnamefont {Kitegi}}, \bibinfo {author} {\bibfnamefont
  {O.}~\bibnamefont {Kononenko}}, \bibinfo {author} {\bibfnamefont
  {M.}~\bibnamefont {Kuntzsch}}, \bibinfo {author} {\bibfnamefont
  {M.}~\bibnamefont {LaBerge}}, \bibinfo {author} {\bibfnamefont
  {S.}~\bibnamefont {Lê}}, \bibinfo {author} {\bibfnamefont {B.}~\bibnamefont
  {Leluan}}, \bibinfo {author} {\bibfnamefont {A.}~\bibnamefont {Loulergue}},
  \bibinfo {author} {\bibfnamefont {V.}~\bibnamefont {Malka}}, \bibinfo
  {author} {\bibfnamefont {F.}~\bibnamefont {Marteau}}, \bibinfo {author}
  {\bibfnamefont {M.~H.~N.}\ \bibnamefont {Guyen}}, \bibinfo {author}
  {\bibfnamefont {D.}~\bibnamefont {Oumbarek-Espinos}}, \bibinfo {author}
  {\bibfnamefont {R.}~\bibnamefont {Pausch}}, \bibinfo {author} {\bibfnamefont
  {D.}~\bibnamefont {Pereira}}, \bibinfo {author} {\bibfnamefont
  {T.}~\bibnamefont {Püschel}}, \bibinfo {author} {\bibfnamefont {J.-P.}\
  \bibnamefont {Ricaud}}, \bibinfo {author} {\bibfnamefont {P.}~\bibnamefont
  {Rommeluere}}, \bibinfo {author} {\bibfnamefont {E.}~\bibnamefont {Roussel}},
  \bibinfo {author} {\bibfnamefont {P.}~\bibnamefont {Rousseau}}, \bibinfo
  {author} {\bibfnamefont {S.}~\bibnamefont {Schöbel}}, \bibinfo {author}
  {\bibfnamefont {M.}~\bibnamefont {Sebdaoui}}, \bibinfo {author}
  {\bibfnamefont {K.}~\bibnamefont {Steiniger}}, \bibinfo {author}
  {\bibfnamefont {K.}~\bibnamefont {Tavakoli}}, \bibinfo {author}
  {\bibfnamefont {C.}~\bibnamefont {Thaury}}, \bibinfo {author} {\bibfnamefont
  {P.}~\bibnamefont {Ufer}}, \bibinfo {author} {\bibfnamefont {M.}~\bibnamefont
  {Valléau}}, \bibinfo {author} {\bibfnamefont {M.}~\bibnamefont
  {Vandenberghe}}, \bibinfo {author} {\bibfnamefont {J.}~\bibnamefont
  {Vétéran}}, \bibinfo {author} {\bibfnamefont {U.}~\bibnamefont {Schramm}},\
  and\ \bibinfo {author} {\bibfnamefont {M.-E.}\ \bibnamefont {Couprie}},\
  }\bibfield  {title} {\bibinfo {title} {Seeded free-electron laser driven by a
  compact laser plasma accelerator},\ }\href
  {https://doi.org/10.1038/s41566-022-01104-w} {\bibfield  {journal} {\bibinfo
  {journal} {Nat. Photonics}\ }\textbf {\bibinfo {volume} {17}},\ \bibinfo
  {pages} {150} (\bibinfo {year} {2023})}\BibitemShut {NoStop}%
\bibitem [{\citenamefont {Ta~Phuoc}\ \emph {et~al.}(2012)\citenamefont
  {Ta~Phuoc}, \citenamefont {Corde}, \citenamefont {Thaury}, \citenamefont
  {Malka}, \citenamefont {Tafzi}, \citenamefont {Goddet}, \citenamefont {Shah},
  \citenamefont {Sebban},\ and\ \citenamefont {Rousse}}]{taphuoc2012}%
  \BibitemOpen
  \bibfield  {author} {\bibinfo {author} {\bibfnamefont {K.}~\bibnamefont
  {Ta~Phuoc}}, \bibinfo {author} {\bibfnamefont {S.}~\bibnamefont {Corde}},
  \bibinfo {author} {\bibfnamefont {C.}~\bibnamefont {Thaury}}, \bibinfo
  {author} {\bibfnamefont {V.}~\bibnamefont {Malka}}, \bibinfo {author}
  {\bibfnamefont {A.}~\bibnamefont {Tafzi}}, \bibinfo {author} {\bibfnamefont
  {J.~P.}\ \bibnamefont {Goddet}}, \bibinfo {author} {\bibfnamefont {R.~C.}\
  \bibnamefont {Shah}}, \bibinfo {author} {\bibfnamefont {S.}~\bibnamefont
  {Sebban}},\ and\ \bibinfo {author} {\bibfnamefont {A.}~\bibnamefont
  {Rousse}},\ }\bibfield  {title} {\bibinfo {title} {All-optical {{Compton}}
  gamma-ray source},\ }\href {https://doi.org/10.1038/nphoton.2012.82}
  {\bibfield  {journal} {\bibinfo  {journal} {Nat. Photonics}\ }\textbf
  {\bibinfo {volume} {6}},\ \bibinfo {pages} {308} (\bibinfo {year}
  {2012})}\BibitemShut {NoStop}%
\bibitem [{\citenamefont {Powers}\ \emph {et~al.}(2014)\citenamefont {Powers},
  \citenamefont {Ghebregziabher}, \citenamefont {Golovin}, \citenamefont {Liu},
  \citenamefont {Chen}, \citenamefont {Banerjee}, \citenamefont {Zhang},\ and\
  \citenamefont {Umstadter}}]{powers2014}%
  \BibitemOpen
  \bibfield  {author} {\bibinfo {author} {\bibfnamefont {N.~D.}\ \bibnamefont
  {Powers}}, \bibinfo {author} {\bibfnamefont {I.}~\bibnamefont
  {Ghebregziabher}}, \bibinfo {author} {\bibfnamefont {G.}~\bibnamefont
  {Golovin}}, \bibinfo {author} {\bibfnamefont {C.}~\bibnamefont {Liu}},
  \bibinfo {author} {\bibfnamefont {S.}~\bibnamefont {Chen}}, \bibinfo {author}
  {\bibfnamefont {S.}~\bibnamefont {Banerjee}}, \bibinfo {author}
  {\bibfnamefont {J.}~\bibnamefont {Zhang}},\ and\ \bibinfo {author}
  {\bibfnamefont {D.~P.}\ \bibnamefont {Umstadter}},\ }\bibfield  {title}
  {\bibinfo {title} {Quasi-monoenergetic and tunable {{X-rays}} from a
  laser-driven {{Compton}} light source},\ }\href
  {https://doi.org/10.1038/nphoton.2013.314} {\bibfield  {journal} {\bibinfo
  {journal} {Nat. Photonics}\ }\textbf {\bibinfo {volume} {8}},\ \bibinfo
  {pages} {28} (\bibinfo {year} {2014})}\BibitemShut {NoStop}%
\bibitem [{\citenamefont {Ma}\ \emph {et~al.}(2023)\citenamefont {Ma},
  \citenamefont {Hua}, \citenamefont {Liu}, \citenamefont {He}, \citenamefont
  {Zhang}, \citenamefont {Chen}, \citenamefont {Yang}, \citenamefont {Ning},
  \citenamefont {Zhang}, \citenamefont {Du},\ and\ \citenamefont
  {Lu}}]{ma2023compact}%
  \BibitemOpen
  \bibfield  {author} {\bibinfo {author} {\bibfnamefont {Y.}~\bibnamefont
  {Ma}}, \bibinfo {author} {\bibfnamefont {J.}~\bibnamefont {Hua}}, \bibinfo
  {author} {\bibfnamefont {D.}~\bibnamefont {Liu}}, \bibinfo {author}
  {\bibfnamefont {Y.}~\bibnamefont {He}}, \bibinfo {author} {\bibfnamefont
  {T.}~\bibnamefont {Zhang}}, \bibinfo {author} {\bibfnamefont
  {J.}~\bibnamefont {Chen}}, \bibinfo {author} {\bibfnamefont {F.}~\bibnamefont
  {Yang}}, \bibinfo {author} {\bibfnamefont {X.}~\bibnamefont {Ning}}, \bibinfo
  {author} {\bibfnamefont {H.}~\bibnamefont {Zhang}}, \bibinfo {author}
  {\bibfnamefont {Y.}~\bibnamefont {Du}},\ and\ \bibinfo {author}
  {\bibfnamefont {W.}~\bibnamefont {Lu}},\ }\bibfield  {title} {\bibinfo
  {title} {Compact {{Polarized X-Ray Source Based}} on {{All-Optical Inverse
  Compton Scattering}}},\ }\href
  {https://doi.org/10.1103/PhysRevApplied.19.014073} {\bibfield  {journal}
  {\bibinfo  {journal} {Phys. Rev. Applied}\ }\textbf {\bibinfo {volume}
  {19}},\ \bibinfo {pages} {014073} (\bibinfo {year} {2023})}\BibitemShut
  {NoStop}%
\bibitem [{\citenamefont {Brümmer}\ \emph {et~al.}(2022)\citenamefont
  {Brümmer}, \citenamefont {Bohlen}, \citenamefont {Grüner}, \citenamefont
  {Osterhoff},\ and\ \citenamefont {Põder}}]{brummer2022}%
  \BibitemOpen
  \bibfield  {author} {\bibinfo {author} {\bibfnamefont {T.}~\bibnamefont
  {Brümmer}}, \bibinfo {author} {\bibfnamefont {S.}~\bibnamefont {Bohlen}},
  \bibinfo {author} {\bibfnamefont {F.}~\bibnamefont {Grüner}}, \bibinfo
  {author} {\bibfnamefont {J.}~\bibnamefont {Osterhoff}},\ and\ \bibinfo
  {author} {\bibfnamefont {K.}~\bibnamefont {Põder}},\ }\bibfield  {title}
  {\bibinfo {title} {Compact all-optical precision-tunable narrowband hard
  {{Compton X-ray}} source},\ }\href
  {https://doi.org/10.1038/s41598-022-20283-8} {\bibfield  {journal} {\bibinfo
  {journal} {Sci. Rep.}\ }\textbf {\bibinfo {volume} {12}},\ \bibinfo {pages}
  {16017} (\bibinfo {year} {2022})}\BibitemShut {NoStop}%
\bibitem [{\citenamefont {Vieira}\ \emph
  {et~al.}(2011{\natexlab{b}})\citenamefont {Vieira}, \citenamefont {Huang},
  \citenamefont {Mori},\ and\ \citenamefont {Silva}}]{vieira2011polarized}%
  \BibitemOpen
  \bibfield  {author} {\bibinfo {author} {\bibfnamefont {J.}~\bibnamefont
  {Vieira}}, \bibinfo {author} {\bibfnamefont {C.-K.}\ \bibnamefont {Huang}},
  \bibinfo {author} {\bibfnamefont {W.~B.}\ \bibnamefont {Mori}},\ and\
  \bibinfo {author} {\bibfnamefont {L.~O.}\ \bibnamefont {Silva}},\ }\bibfield
  {title} {\bibinfo {title} {Polarized beam conditioning in plasma based
  acceleration},\ }\href {https://doi.org/10.1103/PhysRevSTAB.14.071303}
  {\bibfield  {journal} {\bibinfo  {journal} {Phys. Rev. ST Accel. Beams}\
  }\textbf {\bibinfo {volume} {14}},\ \bibinfo {pages} {071303} (\bibinfo
  {year} {2011}{\natexlab{b}})}\BibitemShut {NoStop}%
\bibitem [{\citenamefont {Wen}\ \emph {et~al.}(2019)\citenamefont {Wen},
  \citenamefont {Tamburini},\ and\ \citenamefont {Keitel}}]{wen2019prl}%
  \BibitemOpen
  \bibfield  {author} {\bibinfo {author} {\bibfnamefont {M.}~\bibnamefont
  {Wen}}, \bibinfo {author} {\bibfnamefont {M.}~\bibnamefont {Tamburini}},\
  and\ \bibinfo {author} {\bibfnamefont {C.~H.}\ \bibnamefont {Keitel}},\
  }\bibfield  {title} {\bibinfo {title} {Polarized
  {{Laser-WakeField-Accelerated Kiloampere Electron Beams}}},\ }\href
  {https://doi.org/10.1103/PhysRevLett.122.214801} {\bibfield  {journal}
  {\bibinfo  {journal} {Phys. Rev. Lett.}\ }\textbf {\bibinfo {volume} {122}},\
  \bibinfo {pages} {214801} (\bibinfo {year} {2019})}\BibitemShut {NoStop}%
\bibitem [{\citenamefont {Wu}\ \emph {et~al.}(2019{\natexlab{a}})\citenamefont
  {Wu}, \citenamefont {Ji}, \citenamefont {Geng}, \citenamefont {Yu},
  \citenamefont {Wang}, \citenamefont {Feng}, \citenamefont {Guo},
  \citenamefont {Wang}, \citenamefont {Qin}, \citenamefont {Yan}, \citenamefont
  {Zhang}, \citenamefont {Thomas}, \citenamefont {H{\"u}tzen}, \citenamefont
  {Pukhov}, \citenamefont {B{\"u}scher}, \citenamefont {Shen},\ and\
  \citenamefont {Li}}]{wu2019pre}%
  \BibitemOpen
  \bibfield  {author} {\bibinfo {author} {\bibfnamefont {Y.}~\bibnamefont
  {Wu}}, \bibinfo {author} {\bibfnamefont {L.}~\bibnamefont {Ji}}, \bibinfo
  {author} {\bibfnamefont {X.}~\bibnamefont {Geng}}, \bibinfo {author}
  {\bibfnamefont {Q.}~\bibnamefont {Yu}}, \bibinfo {author} {\bibfnamefont
  {N.}~\bibnamefont {Wang}}, \bibinfo {author} {\bibfnamefont {B.}~\bibnamefont
  {Feng}}, \bibinfo {author} {\bibfnamefont {Z.}~\bibnamefont {Guo}}, \bibinfo
  {author} {\bibfnamefont {W.}~\bibnamefont {Wang}}, \bibinfo {author}
  {\bibfnamefont {C.}~\bibnamefont {Qin}}, \bibinfo {author} {\bibfnamefont
  {X.}~\bibnamefont {Yan}}, \bibinfo {author} {\bibfnamefont {L.}~\bibnamefont
  {Zhang}}, \bibinfo {author} {\bibfnamefont {J.}~\bibnamefont {Thomas}},
  \bibinfo {author} {\bibfnamefont {A.}~\bibnamefont {H{\"u}tzen}}, \bibinfo
  {author} {\bibfnamefont {A.}~\bibnamefont {Pukhov}}, \bibinfo {author}
  {\bibfnamefont {M.}~\bibnamefont {B{\"u}scher}}, \bibinfo {author}
  {\bibfnamefont {B.}~\bibnamefont {Shen}},\ and\ \bibinfo {author}
  {\bibfnamefont {R.}~\bibnamefont {Li}},\ }\bibfield  {title} {\bibinfo
  {title} {Polarized electron acceleration in beam-driven plasma wakefield
  based on density down-ramp injection},\ }\href
  {https://doi.org/10.1103/PhysRevE.100.043202} {\bibfield  {journal} {\bibinfo
   {journal} {Phys. Rev. E}\ }\textbf {\bibinfo {volume} {100}},\ \bibinfo
  {pages} {043202} (\bibinfo {year} {2019}{\natexlab{a}})}\BibitemShut
  {NoStop}%
\bibitem [{\citenamefont {Wu}\ \emph {et~al.}(2019{\natexlab{b}})\citenamefont
  {Wu}, \citenamefont {Ji}, \citenamefont {Geng}, \citenamefont {Yu},
  \citenamefont {Wang}, \citenamefont {Feng}, \citenamefont {Guo},
  \citenamefont {Wang}, \citenamefont {Qin}, \citenamefont {Yan}, \citenamefont
  {Zhang}, \citenamefont {Thomas}, \citenamefont {H{\"u}tzen}, \citenamefont
  {B{\"u}scher}, \citenamefont {Rakitzis}, \citenamefont {Pukhov},
  \citenamefont {Shen},\ and\ \citenamefont {Li}}]{wu2019njp}%
  \BibitemOpen
  \bibfield  {author} {\bibinfo {author} {\bibfnamefont {Y.}~\bibnamefont
  {Wu}}, \bibinfo {author} {\bibfnamefont {L.}~\bibnamefont {Ji}}, \bibinfo
  {author} {\bibfnamefont {X.}~\bibnamefont {Geng}}, \bibinfo {author}
  {\bibfnamefont {Q.}~\bibnamefont {Yu}}, \bibinfo {author} {\bibfnamefont
  {N.}~\bibnamefont {Wang}}, \bibinfo {author} {\bibfnamefont {B.}~\bibnamefont
  {Feng}}, \bibinfo {author} {\bibfnamefont {Z.}~\bibnamefont {Guo}}, \bibinfo
  {author} {\bibfnamefont {W.}~\bibnamefont {Wang}}, \bibinfo {author}
  {\bibfnamefont {C.}~\bibnamefont {Qin}}, \bibinfo {author} {\bibfnamefont
  {X.}~\bibnamefont {Yan}}, \bibinfo {author} {\bibfnamefont {L.}~\bibnamefont
  {Zhang}}, \bibinfo {author} {\bibfnamefont {J.}~\bibnamefont {Thomas}},
  \bibinfo {author} {\bibfnamefont {A.}~\bibnamefont {H{\"u}tzen}}, \bibinfo
  {author} {\bibfnamefont {M.}~\bibnamefont {B{\"u}scher}}, \bibinfo {author}
  {\bibfnamefont {T.~P.}\ \bibnamefont {Rakitzis}}, \bibinfo {author}
  {\bibfnamefont {A.}~\bibnamefont {Pukhov}}, \bibinfo {author} {\bibfnamefont
  {B.}~\bibnamefont {Shen}},\ and\ \bibinfo {author} {\bibfnamefont
  {R.}~\bibnamefont {Li}},\ }\bibfield  {title} {\bibinfo {title} {Polarized
  electron-beam acceleration driven by vortex laser pulses},\ }\href
  {https://doi.org/10.1088/1367-2630/ab2fd7} {\bibfield  {journal} {\bibinfo
  {journal} {New J. Phys.}\ }\textbf {\bibinfo {volume} {21}},\ \bibinfo
  {pages} {073052} (\bibinfo {year} {2019}{\natexlab{b}})}\BibitemShut
  {NoStop}%
\bibitem [{\citenamefont {Nie}\ \emph {et~al.}(2021)\citenamefont {Nie},
  \citenamefont {Li}, \citenamefont {Morales}, \citenamefont {Patchkovskii},
  \citenamefont {Smirnova}, \citenamefont {An}, \citenamefont {Nambu},
  \citenamefont {Matteo}, \citenamefont {Marsh}, \citenamefont {Tsung},
  \citenamefont {Mori},\ and\ \citenamefont {Joshi}}]{nie2021}%
  \BibitemOpen
  \bibfield  {author} {\bibinfo {author} {\bibfnamefont {Z.}~\bibnamefont
  {Nie}}, \bibinfo {author} {\bibfnamefont {F.}~\bibnamefont {Li}}, \bibinfo
  {author} {\bibfnamefont {F.}~\bibnamefont {Morales}}, \bibinfo {author}
  {\bibfnamefont {S.}~\bibnamefont {Patchkovskii}}, \bibinfo {author}
  {\bibfnamefont {O.}~\bibnamefont {Smirnova}}, \bibinfo {author}
  {\bibfnamefont {W.}~\bibnamefont {An}}, \bibinfo {author} {\bibfnamefont
  {N.}~\bibnamefont {Nambu}}, \bibinfo {author} {\bibfnamefont
  {D.}~\bibnamefont {Matteo}}, \bibinfo {author} {\bibfnamefont {K.~A.}\
  \bibnamefont {Marsh}}, \bibinfo {author} {\bibfnamefont {F.}~\bibnamefont
  {Tsung}}, \bibinfo {author} {\bibfnamefont {W.~B.}\ \bibnamefont {Mori}},\
  and\ \bibinfo {author} {\bibfnamefont {C.}~\bibnamefont {Joshi}},\ }\bibfield
   {title} {\bibinfo {title} {{\emph{In }}{{{\emph{Situ}}}} {{Generation}} of
  {{High-Energy Spin-Polarized Electrons}} in a {{Beam-Driven Plasma Wakefield
  Accelerator}}},\ }\href {https://doi.org/10.1103/PhysRevLett.126.054801}
  {\bibfield  {journal} {\bibinfo  {journal} {Phys. Rev. Lett.}\ }\textbf
  {\bibinfo {volume} {126}},\ \bibinfo {pages} {054801} (\bibinfo {year}
  {2021})}\BibitemShut {NoStop}%
\bibitem [{\citenamefont {Fan}\ \emph {et~al.}(2022)\citenamefont {Fan},
  \citenamefont {Liu}, \citenamefont {Li}, \citenamefont {Qu}, \citenamefont
  {Yu}, \citenamefont {Kong}, \citenamefont {Weng}, \citenamefont {Chen},
  \citenamefont {B{\"u}scher}, \citenamefont {Gibbon}, \citenamefont {Kawata},\
  and\ \citenamefont {Sheng}}]{fan2022}%
  \BibitemOpen
  \bibfield  {author} {\bibinfo {author} {\bibfnamefont {H.~C.}\ \bibnamefont
  {Fan}}, \bibinfo {author} {\bibfnamefont {X.~Y.}\ \bibnamefont {Liu}},
  \bibinfo {author} {\bibfnamefont {X.~F.}\ \bibnamefont {Li}}, \bibinfo
  {author} {\bibfnamefont {J.~F.}\ \bibnamefont {Qu}}, \bibinfo {author}
  {\bibfnamefont {Q.}~\bibnamefont {Yu}}, \bibinfo {author} {\bibfnamefont
  {Q.}~\bibnamefont {Kong}}, \bibinfo {author} {\bibfnamefont {S.~M.}\
  \bibnamefont {Weng}}, \bibinfo {author} {\bibfnamefont {M.}~\bibnamefont
  {Chen}}, \bibinfo {author} {\bibfnamefont {M.}~\bibnamefont {B{\"u}scher}},
  \bibinfo {author} {\bibfnamefont {P.}~\bibnamefont {Gibbon}}, \bibinfo
  {author} {\bibfnamefont {S.}~\bibnamefont {Kawata}},\ and\ \bibinfo {author}
  {\bibfnamefont {Z.~M.}\ \bibnamefont {Sheng}},\ }\bibfield  {title} {\bibinfo
  {title} {Control of electron beam polarization in the bubble regime of
  laser-wakefield acceleration},\ }\href
  {https://doi.org/10.1088/1367-2630/ac8951} {\bibfield  {journal} {\bibinfo
  {journal} {New J. Phys.}\ }\textbf {\bibinfo {volume} {24}},\ \bibinfo
  {pages} {083047} (\bibinfo {year} {2022})}\BibitemShut {NoStop}%
\bibitem [{\citenamefont {Rakitzis}\ \emph {et~al.}(2003)\citenamefont
  {Rakitzis}, \citenamefont {Samartzis}, \citenamefont {Toomes}, \citenamefont
  {Kitsopoulos}, \citenamefont {Brown}, \citenamefont {{Balint-Kurti}},
  \citenamefont {Vasyutinskii},\ and\ \citenamefont {Beswick}}]{rakitzis2003}%
  \BibitemOpen
  \bibfield  {author} {\bibinfo {author} {\bibfnamefont {T.~P.}\ \bibnamefont
  {Rakitzis}}, \bibinfo {author} {\bibfnamefont {P.~C.}\ \bibnamefont
  {Samartzis}}, \bibinfo {author} {\bibfnamefont {R.~L.}\ \bibnamefont
  {Toomes}}, \bibinfo {author} {\bibfnamefont {T.~N.}\ \bibnamefont
  {Kitsopoulos}}, \bibinfo {author} {\bibfnamefont {A.}~\bibnamefont {Brown}},
  \bibinfo {author} {\bibfnamefont {G.~G.}\ \bibnamefont {{Balint-Kurti}}},
  \bibinfo {author} {\bibfnamefont {O.~S.}\ \bibnamefont {Vasyutinskii}},\ and\
  \bibinfo {author} {\bibfnamefont {J.~A.}\ \bibnamefont {Beswick}},\
  }\bibfield  {title} {\bibinfo {title} {Spin-{{Polarized Hydrogen Atoms}} from
  {{Molecular Photodissociation}}},\ }\href@noop {} {\bibfield  {journal}
  {\bibinfo  {journal} {Science}\ }\textbf {\bibinfo {volume} {300}},\ \bibinfo
  {pages} {1936} (\bibinfo {year} {2003})}\BibitemShut {NoStop}%
\bibitem [{\citenamefont {Sofikitis}\ \emph {et~al.}(2008)\citenamefont
  {Sofikitis}, \citenamefont {Rubio-Lago}, \citenamefont {Bougas},
  \citenamefont {Alexander},\ and\ \citenamefont {Rakitzis}}]{sofikitis2008}%
  \BibitemOpen
  \bibfield  {author} {\bibinfo {author} {\bibfnamefont {D.}~\bibnamefont
  {Sofikitis}}, \bibinfo {author} {\bibfnamefont {L.}~\bibnamefont
  {Rubio-Lago}}, \bibinfo {author} {\bibfnamefont {L.}~\bibnamefont {Bougas}},
  \bibinfo {author} {\bibfnamefont {A.~J.}\ \bibnamefont {Alexander}},\ and\
  \bibinfo {author} {\bibfnamefont {T.~P.}\ \bibnamefont {Rakitzis}},\
  }\bibfield  {title} {\bibinfo {title} {Laser detection of spin-polarized
  hydrogen from {{HCl}} and {{HBr}} photodissociation: {{Comparison}} of {{H-}}
  and halogen-atom polarizations},\ }\href {https://doi.org/10.1063/1.2989803}
  {\bibfield  {journal} {\bibinfo  {journal} {J. Chem. Phys.}\ }\textbf
  {\bibinfo {volume} {129}},\ \bibinfo {pages} {144302} (\bibinfo {year}
  {2008})}\BibitemShut {NoStop}%
\bibitem [{\citenamefont {Sofikitis}\ \emph {et~al.}(2018)\citenamefont
  {Sofikitis}, \citenamefont {Kannis}, \citenamefont {Boulogiannis},\ and\
  \citenamefont {Rakitzis}}]{sofikitis2018}%
  \BibitemOpen
  \bibfield  {author} {\bibinfo {author} {\bibfnamefont {D.}~\bibnamefont
  {Sofikitis}}, \bibinfo {author} {\bibfnamefont {C.~S.}\ \bibnamefont
  {Kannis}}, \bibinfo {author} {\bibfnamefont {G.~K.}\ \bibnamefont
  {Boulogiannis}},\ and\ \bibinfo {author} {\bibfnamefont {T.~P.}\ \bibnamefont
  {Rakitzis}},\ }\bibfield  {title} {\bibinfo {title} {Ultrahigh-{{Density
  Spin-Polarized H}} and {{D Observed}} via {{Magnetization Quantum Beats}}},\
  }\href {https://doi.org/10.1103/PhysRevLett.121.083001} {\bibfield  {journal}
  {\bibinfo  {journal} {Phys. Rev. Lett.}\ }\textbf {\bibinfo {volume} {121}},\
  \bibinfo {pages} {083001} (\bibinfo {year} {2018})}\BibitemShut {NoStop}%
\bibitem [{\citenamefont {Spiliotis}\ \emph {et~al.}(2021)\citenamefont
  {Spiliotis}, \citenamefont {Xygkis}, \citenamefont {Koutrakis}, \citenamefont
  {Sofikitis},\ and\ \citenamefont {Rakitzis}}]{spiliotis2021}%
  \BibitemOpen
  \bibfield  {author} {\bibinfo {author} {\bibfnamefont {A.~K.}\ \bibnamefont
  {Spiliotis}}, \bibinfo {author} {\bibfnamefont {M.}~\bibnamefont {Xygkis}},
  \bibinfo {author} {\bibfnamefont {M.~E.}\ \bibnamefont {Koutrakis}}, \bibinfo
  {author} {\bibfnamefont {D.}~\bibnamefont {Sofikitis}},\ and\ \bibinfo
  {author} {\bibfnamefont {T.~P.}\ \bibnamefont {Rakitzis}},\ }\bibfield
  {title} {\bibinfo {title} {Depolarization of spin-polarized hydrogen via
  collisions with chlorine atoms at ultrahigh density},\ }\href
  {https://doi.org/10.1016/j.chphi.2021.100022} {\bibfield  {journal} {\bibinfo
   {journal} {Chem. Phys. Impact}\ }\textbf {\bibinfo {volume} {2}},\ \bibinfo
  {pages} {100022} (\bibinfo {year} {2021})}\BibitemShut {NoStop}%
\bibitem [{\citenamefont {Sun}\ \emph {et~al.}(2022)\citenamefont {Sun},
  \citenamefont {Zhao}, \citenamefont {Xue}, \citenamefont {Lu}, \citenamefont
  {Ji}, \citenamefont {Wan}, \citenamefont {Wang}, \citenamefont {Salamin},\
  and\ \citenamefont {Li}}]{sun2022}%
  \BibitemOpen
  \bibfield  {author} {\bibinfo {author} {\bibfnamefont {T.}~\bibnamefont
  {Sun}}, \bibinfo {author} {\bibfnamefont {Q.}~\bibnamefont {Zhao}}, \bibinfo
  {author} {\bibfnamefont {K.}~\bibnamefont {Xue}}, \bibinfo {author}
  {\bibfnamefont {Z.-W.}\ \bibnamefont {Lu}}, \bibinfo {author} {\bibfnamefont
  {L.-L.}\ \bibnamefont {Ji}}, \bibinfo {author} {\bibfnamefont
  {F.}~\bibnamefont {Wan}}, \bibinfo {author} {\bibfnamefont {Y.}~\bibnamefont
  {Wang}}, \bibinfo {author} {\bibfnamefont {Y.~I.}\ \bibnamefont {Salamin}},\
  and\ \bibinfo {author} {\bibfnamefont {J.-X.}\ \bibnamefont {Li}},\
  }\bibfield  {title} {\bibinfo {title} {Production of polarized particle beams
  via ultraintense laser pulses},\ }\href
  {https://doi.org/10.1007/s41614-022-00099-9} {\bibfield  {journal} {\bibinfo
  {journal} {Rev. Mod. Plasma Phys.}\ }\textbf {\bibinfo {volume} {6}},\
  \bibinfo {pages} {38} (\bibinfo {year} {2022})}\BibitemShut {NoStop}%
\bibitem [{\citenamefont {Gay}(2009)}]{gay2009}%
  \BibitemOpen
  \bibfield  {author} {\bibinfo {author} {\bibfnamefont {T.~J.}\ \bibnamefont
  {Gay}},\ }\bibfield  {title} {\bibinfo {title} {Physics and {{Technology}} of
  {{Polarized Electron Scattering}} from {{Atoms}} and {{Molecules}}},\ }in\
  \href {https://doi.org/10.1016/S1049-250X(09)57004-8} {\emph {\bibinfo
  {booktitle} {Advances in {{Atomic}}, {{Molecular}}, and {{Optical Physics}},
  {{Vol}} 57}}},\ Vol.~\bibinfo {volume} {57},\ \bibinfo {editor} {edited by\
  \bibinfo {editor} {\bibfnamefont {E.}~\bibnamefont {Arimondo}}, \bibinfo
  {editor} {\bibfnamefont {P.~R.}\ \bibnamefont {Berman}},\ and\ \bibinfo
  {editor} {\bibfnamefont {C.~C.}\ \bibnamefont {Lin}}}\ (\bibinfo  {publisher}
  {{Elsevier Academic Press Inc}},\ \bibinfo {address} {{San Diego}},\ \bibinfo
  {year} {2009})\ pp.\ \bibinfo {pages} {157--247}\BibitemShut {NoStop}%
\bibitem [{\citenamefont {Abe}\ \emph {et~al.}(1995)\citenamefont {Abe} \emph
  {et~al.}}]{abe1995}%
  \BibitemOpen
  \bibfield  {author} {\bibinfo {author} {\bibfnamefont {K.}~\bibnamefont
  {Abe}} \emph {et~al.} (\bibinfo {collaboration} {E143 Collaboration}),\
  }\bibfield  {title} {\bibinfo {title} {Precision measurement of the deuteron
  spin structure function ${g}_{1}^{\mathit{d}}$},\ }\href
  {https://doi.org/10.1103/PhysRevLett.75.25} {\bibfield  {journal} {\bibinfo
  {journal} {Phys. Rev. Lett.}\ }\textbf {\bibinfo {volume} {75}},\ \bibinfo
  {pages} {25} (\bibinfo {year} {1995})}\BibitemShut {NoStop}%
\bibitem [{\citenamefont {Alexakhin}\ \emph {et~al.}(2007)\citenamefont
  {Alexakhin}, \citenamefont {Alexandrov}, \citenamefont {Alexeev},
  \citenamefont {Alexeev}, \citenamefont {Amoroso}, \citenamefont {Bade{\l}ek},
  \citenamefont {Balestra}, \citenamefont {Ball}, \citenamefont {Barth},
  \citenamefont {Baum} \emph {et~al.}}]{alexakhin2007}%
  \BibitemOpen
  \bibfield  {author} {\bibinfo {author} {\bibfnamefont {V.~Y.}\ \bibnamefont
  {Alexakhin}}, \bibinfo {author} {\bibfnamefont {Y.}~\bibnamefont
  {Alexandrov}}, \bibinfo {author} {\bibfnamefont {G.~D.}\ \bibnamefont
  {Alexeev}}, \bibinfo {author} {\bibfnamefont {M.}~\bibnamefont {Alexeev}},
  \bibinfo {author} {\bibfnamefont {A.}~\bibnamefont {Amoroso}}, \bibinfo
  {author} {\bibfnamefont {B.}~\bibnamefont {Bade{\l}ek}}, \bibinfo {author}
  {\bibfnamefont {F.}~\bibnamefont {Balestra}}, \bibinfo {author}
  {\bibfnamefont {J.}~\bibnamefont {Ball}}, \bibinfo {author} {\bibfnamefont
  {J.}~\bibnamefont {Barth}}, \bibinfo {author} {\bibfnamefont
  {G.}~\bibnamefont {Baum}}, \emph {et~al.},\ }\bibfield  {title} {\bibinfo
  {title} {The deuteron spin-dependent structure function
  ${g}_{1}^{\mathit{d}}$ and its first moment},\ }\href@noop {} {\bibfield
  {journal} {\bibinfo  {journal} {Phys. Lett. B}\ }\textbf {\bibinfo {volume}
  {647}},\ \bibinfo {pages} {8} (\bibinfo {year} {2007})}\BibitemShut {NoStop}%
\bibitem [{\citenamefont {Li}\ \emph {et~al.}(2020)\citenamefont {Li},
  \citenamefont {Shaisultanov}, \citenamefont {Chen}, \citenamefont {Wan},
  \citenamefont {Hatsagortsyan}, \citenamefont {Keitel},\ and\ \citenamefont
  {Li}}]{li2020}%
  \BibitemOpen
  \bibfield  {author} {\bibinfo {author} {\bibfnamefont {Y.-F.}\ \bibnamefont
  {Li}}, \bibinfo {author} {\bibfnamefont {R.}~\bibnamefont {Shaisultanov}},
  \bibinfo {author} {\bibfnamefont {Y.-Y.}\ \bibnamefont {Chen}}, \bibinfo
  {author} {\bibfnamefont {F.}~\bibnamefont {Wan}}, \bibinfo {author}
  {\bibfnamefont {K.~Z.}\ \bibnamefont {Hatsagortsyan}}, \bibinfo {author}
  {\bibfnamefont {C.~H.}\ \bibnamefont {Keitel}},\ and\ \bibinfo {author}
  {\bibfnamefont {J.-X.}\ \bibnamefont {Li}},\ }\bibfield  {title} {\bibinfo
  {title} {Polarized {{Ultrashort Brilliant Multi-GeV}} {$\gamma$} {{Rays}} via
  {{Single-Shot Laser-Electron Interaction}}},\ }\href
  {https://doi.org/10.1103/PhysRevLett.124.014801} {\bibfield  {journal}
  {\bibinfo  {journal} {Phys. Rev. Lett.}\ }\textbf {\bibinfo {volume} {124}},\
  \bibinfo {pages} {014801} (\bibinfo {year} {2020})}\BibitemShut {NoStop}%
\bibitem [{\citenamefont {Wang}\ \emph {et~al.}(2022)\citenamefont {Wang},
  \citenamefont {Ababekri}, \citenamefont {Wan}, \citenamefont {Zhao},
  \citenamefont {Lv}, \citenamefont {Ren}, \citenamefont {Xu}, \citenamefont
  {Zhao},\ and\ \citenamefont {Li}}]{wang2022}%
  \BibitemOpen
  \bibfield  {author} {\bibinfo {author} {\bibfnamefont {Y.}~\bibnamefont
  {Wang}}, \bibinfo {author} {\bibfnamefont {M.}~\bibnamefont {Ababekri}},
  \bibinfo {author} {\bibfnamefont {F.}~\bibnamefont {Wan}}, \bibinfo {author}
  {\bibfnamefont {Q.}~\bibnamefont {Zhao}}, \bibinfo {author} {\bibfnamefont
  {C.}~\bibnamefont {Lv}}, \bibinfo {author} {\bibfnamefont {X.-G.}\
  \bibnamefont {Ren}}, \bibinfo {author} {\bibfnamefont {Z.-F.}\ \bibnamefont
  {Xu}}, \bibinfo {author} {\bibfnamefont {Y.-T.}\ \bibnamefont {Zhao}},\ and\
  \bibinfo {author} {\bibfnamefont {J.-X.}\ \bibnamefont {Li}},\ }\bibfield
  {title} {\bibinfo {title} {Brilliant circularly polarized {$\gamma$}-ray
  sources via single-shot laser plasma interaction},\ }\href
  {https://doi.org/10.1364/OL.462612} {\bibfield  {journal} {\bibinfo
  {journal} {Opt. Lett.}\ }\textbf {\bibinfo {volume} {47}},\ \bibinfo {pages}
  {3355} (\bibinfo {year} {2022})}\BibitemShut {NoStop}%
\bibitem [{\citenamefont {Barth}\ and\ \citenamefont
  {Smirnova}(2013)}]{barth2013}%
  \BibitemOpen
  \bibfield  {author} {\bibinfo {author} {\bibfnamefont {I.}~\bibnamefont
  {Barth}}\ and\ \bibinfo {author} {\bibfnamefont {O.}~\bibnamefont
  {Smirnova}},\ }\bibfield  {title} {\bibinfo {title} {Spin-polarized electrons
  produced by strong-field ionization},\ }\href
  {https://doi.org/10.1103/PhysRevA.88.013401} {\bibfield  {journal} {\bibinfo
  {journal} {Phys. Rev. A}\ }\textbf {\bibinfo {volume} {88}},\ \bibinfo
  {pages} {013401} (\bibinfo {year} {2013})}\BibitemShut {NoStop}%
\bibitem [{\citenamefont {Shaw}\ \emph {et~al.}(2017)\citenamefont {Shaw},
  \citenamefont {Lemos}, \citenamefont {Amorim}, \citenamefont
  {{Vafaei-Najafabadi}}, \citenamefont {Marsh}, \citenamefont {Tsung},
  \citenamefont {Mori},\ and\ \citenamefont {Joshi}}]{shaw2017}%
  \BibitemOpen
  \bibfield  {author} {\bibinfo {author} {\bibfnamefont {J.~L.}\ \bibnamefont
  {Shaw}}, \bibinfo {author} {\bibfnamefont {N.}~\bibnamefont {Lemos}},
  \bibinfo {author} {\bibfnamefont {L.~D.}\ \bibnamefont {Amorim}}, \bibinfo
  {author} {\bibfnamefont {N.}~\bibnamefont {{Vafaei-Najafabadi}}}, \bibinfo
  {author} {\bibfnamefont {K.~A.}\ \bibnamefont {Marsh}}, \bibinfo {author}
  {\bibfnamefont {F.~S.}\ \bibnamefont {Tsung}}, \bibinfo {author}
  {\bibfnamefont {W.~B.}\ \bibnamefont {Mori}},\ and\ \bibinfo {author}
  {\bibfnamefont {C.}~\bibnamefont {Joshi}},\ }\bibfield  {title} {\bibinfo
  {title} {Role of {{Direct Laser Acceleration}} of {{Electrons}} in a {{Laser
  Wakefield Accelerator}} with {{Ionization Injection}}},\ }\href
  {https://doi.org/10.1103/PhysRevLett.118.064801} {\bibfield  {journal}
  {\bibinfo  {journal} {Phys. Rev. Lett.}\ }\textbf {\bibinfo {volume} {118}},\
  \bibinfo {pages} {064801} (\bibinfo {year} {2017})}\BibitemShut {NoStop}%
\bibitem [{\citenamefont {Salamin}(2006)}]{salamin2006njp}%
  \BibitemOpen
  \bibfield  {author} {\bibinfo {author} {\bibfnamefont {Y.~I.}\ \bibnamefont
  {Salamin}},\ }\bibfield  {title} {\bibinfo {title} {Accurate fields of a
  radially polarized {{Gaussian}} laser beam},\ }\href
  {https://doi.org/10.1088/1367-2630/8/8/133} {\bibfield  {journal} {\bibinfo
  {journal} {New J. Phys.}\ }\textbf {\bibinfo {volume} {8}},\ \bibinfo {pages}
  {133} (\bibinfo {year} {2006})}\BibitemShut {NoStop}%
\bibitem [{\citenamefont {Varin}\ \emph {et~al.}(2013)\citenamefont {Varin},
  \citenamefont {Payeur}, \citenamefont {Marceau}, \citenamefont {Fourmaux},
  \citenamefont {April}, \citenamefont {Schmidt}, \citenamefont {Fortin},
  \citenamefont {Thir{\'e}}, \citenamefont {Brabec}, \citenamefont
  {L{\'e}gar{\'e}}, \citenamefont {Kieffer},\ and\ \citenamefont
  {Pich{\'e}}}]{varin2013}%
  \BibitemOpen
  \bibfield  {author} {\bibinfo {author} {\bibfnamefont {C.}~\bibnamefont
  {Varin}}, \bibinfo {author} {\bibfnamefont {S.}~\bibnamefont {Payeur}},
  \bibinfo {author} {\bibfnamefont {V.}~\bibnamefont {Marceau}}, \bibinfo
  {author} {\bibfnamefont {S.}~\bibnamefont {Fourmaux}}, \bibinfo {author}
  {\bibfnamefont {A.}~\bibnamefont {April}}, \bibinfo {author} {\bibfnamefont
  {B.}~\bibnamefont {Schmidt}}, \bibinfo {author} {\bibfnamefont {P.-L.}\
  \bibnamefont {Fortin}}, \bibinfo {author} {\bibfnamefont {N.}~\bibnamefont
  {Thir{\'e}}}, \bibinfo {author} {\bibfnamefont {T.}~\bibnamefont {Brabec}},
  \bibinfo {author} {\bibfnamefont {F.}~\bibnamefont {L{\'e}gar{\'e}}},
  \bibinfo {author} {\bibfnamefont {J.-C.}\ \bibnamefont {Kieffer}},\ and\
  \bibinfo {author} {\bibfnamefont {M.}~\bibnamefont {Pich{\'e}}},\ }\bibfield
  {title} {\bibinfo {title} {Direct {{Electron Acceleration}} with {{Radially
  Polarized Laser Beams}}},\ }\href {https://doi.org/10.3390/app3010070}
  {\bibfield  {journal} {\bibinfo  {journal} {Appl. Sci.}\ }\textbf {\bibinfo
  {volume} {3}},\ \bibinfo {pages} {70} (\bibinfo {year} {2013})}\BibitemShut
  {NoStop}%
\bibitem [{\citenamefont {Tsakiris}\ \emph {et~al.}(2000)\citenamefont
  {Tsakiris}, \citenamefont {Gahn},\ and\ \citenamefont
  {Tripathi}}]{tsakiris2000}%
  \BibitemOpen
  \bibfield  {author} {\bibinfo {author} {\bibfnamefont {G.~D.}\ \bibnamefont
  {Tsakiris}}, \bibinfo {author} {\bibfnamefont {C.}~\bibnamefont {Gahn}},\
  and\ \bibinfo {author} {\bibfnamefont {V.~K.}\ \bibnamefont {Tripathi}},\
  }\bibfield  {title} {\bibinfo {title} {Laser induced electron acceleration in
  the presence of static electric and magnetic fields in a plasma},\ }\href
  {https://doi.org/10.1063/1.874154} {\bibfield  {journal} {\bibinfo  {journal}
  {Phys. Plasmas}\ }\textbf {\bibinfo {volume} {7}},\ \bibinfo {pages} {3017}
  (\bibinfo {year} {2000})}\BibitemShut {NoStop}%
\bibitem [{\citenamefont {Schmid}\ \emph {et~al.}(2010)\citenamefont {Schmid},
  \citenamefont {Buck}, \citenamefont {Sears}, \citenamefont {Mikhailova},
  \citenamefont {Tautz}, \citenamefont {Herrmann}, \citenamefont {Geissler},
  \citenamefont {Krausz},\ and\ \citenamefont {Veisz}}]{schmid2010}%
  \BibitemOpen
  \bibfield  {author} {\bibinfo {author} {\bibfnamefont {K.}~\bibnamefont
  {Schmid}}, \bibinfo {author} {\bibfnamefont {A.}~\bibnamefont {Buck}},
  \bibinfo {author} {\bibfnamefont {C.~M.~S.}\ \bibnamefont {Sears}}, \bibinfo
  {author} {\bibfnamefont {J.~M.}\ \bibnamefont {Mikhailova}}, \bibinfo
  {author} {\bibfnamefont {R.}~\bibnamefont {Tautz}}, \bibinfo {author}
  {\bibfnamefont {D.}~\bibnamefont {Herrmann}}, \bibinfo {author}
  {\bibfnamefont {M.}~\bibnamefont {Geissler}}, \bibinfo {author}
  {\bibfnamefont {F.}~\bibnamefont {Krausz}},\ and\ \bibinfo {author}
  {\bibfnamefont {L.}~\bibnamefont {Veisz}},\ }\bibfield  {title} {\bibinfo
  {title} {Density-transition based electron injector for laser driven
  wakefield accelerators},\ }\href
  {https://doi.org/10.1103/PhysRevSTAB.13.091301} {\bibfield  {journal}
  {\bibinfo  {journal} {Phys. Rev. ST Accel. Beams}\ }\textbf {\bibinfo
  {volume} {13}},\ \bibinfo {pages} {091301} (\bibinfo {year}
  {2010})}\BibitemShut {NoStop}%
\bibitem [{\citenamefont {Buck}\ \emph {et~al.}(2013)\citenamefont {Buck},
  \citenamefont {Wenz}, \citenamefont {Xu}, \citenamefont {Khrennikov},
  \citenamefont {Schmid}, \citenamefont {Heigoldt}, \citenamefont {Mikhailova},
  \citenamefont {Geissler}, \citenamefont {Shen}, \citenamefont {Krausz},
  \citenamefont {Karsch},\ and\ \citenamefont {Veisz}}]{buck2013}%
  \BibitemOpen
  \bibfield  {author} {\bibinfo {author} {\bibfnamefont {A.}~\bibnamefont
  {Buck}}, \bibinfo {author} {\bibfnamefont {J.}~\bibnamefont {Wenz}}, \bibinfo
  {author} {\bibfnamefont {J.}~\bibnamefont {Xu}}, \bibinfo {author}
  {\bibfnamefont {K.}~\bibnamefont {Khrennikov}}, \bibinfo {author}
  {\bibfnamefont {K.}~\bibnamefont {Schmid}}, \bibinfo {author} {\bibfnamefont
  {M.}~\bibnamefont {Heigoldt}}, \bibinfo {author} {\bibfnamefont {J.~M.}\
  \bibnamefont {Mikhailova}}, \bibinfo {author} {\bibfnamefont
  {M.}~\bibnamefont {Geissler}}, \bibinfo {author} {\bibfnamefont
  {B.}~\bibnamefont {Shen}}, \bibinfo {author} {\bibfnamefont {F.}~\bibnamefont
  {Krausz}}, \bibinfo {author} {\bibfnamefont {S.}~\bibnamefont {Karsch}},\
  and\ \bibinfo {author} {\bibfnamefont {L.}~\bibnamefont {Veisz}},\ }\bibfield
   {title} {\bibinfo {title} {Shock-{{Front Injector}} for {{High-Quality
  Laser-Plasma Acceleration}}},\ }\href
  {https://doi.org/10.1103/PhysRevLett.110.185006} {\bibfield  {journal}
  {\bibinfo  {journal} {Phys. Rev. Lett.}\ }\textbf {\bibinfo {volume} {110}},\
  \bibinfo {pages} {185006} (\bibinfo {year} {2013})}\BibitemShut {NoStop}%
\bibitem [{\citenamefont {Chou}\ \emph {et~al.}(2016)\citenamefont {Chou},
  \citenamefont {Xu}, \citenamefont {Khrennikov}, \citenamefont {Cardenas},
  \citenamefont {Wenz}, \citenamefont {Heigoldt}, \citenamefont {Hofmann},
  \citenamefont {Veisz},\ and\ \citenamefont {Karsch}}]{chou2016}%
  \BibitemOpen
  \bibfield  {author} {\bibinfo {author} {\bibfnamefont {S.}~\bibnamefont
  {Chou}}, \bibinfo {author} {\bibfnamefont {J.}~\bibnamefont {Xu}}, \bibinfo
  {author} {\bibfnamefont {K.}~\bibnamefont {Khrennikov}}, \bibinfo {author}
  {\bibfnamefont {D.~E.}\ \bibnamefont {Cardenas}}, \bibinfo {author}
  {\bibfnamefont {J.}~\bibnamefont {Wenz}}, \bibinfo {author} {\bibfnamefont
  {M.}~\bibnamefont {Heigoldt}}, \bibinfo {author} {\bibfnamefont
  {L.}~\bibnamefont {Hofmann}}, \bibinfo {author} {\bibfnamefont
  {L.}~\bibnamefont {Veisz}},\ and\ \bibinfo {author} {\bibfnamefont
  {S.}~\bibnamefont {Karsch}},\ }\bibfield  {title} {\bibinfo {title}
  {Collective {{Deceleration}} of {{Laser-Driven Electron Bunches}}},\ }\href
  {https://doi.org/10.1103/PhysRevLett.117.144801} {\bibfield  {journal}
  {\bibinfo  {journal} {Phys. Rev. Lett.}\ }\textbf {\bibinfo {volume} {117}},\
  \bibinfo {pages} {144801} (\bibinfo {year} {2016})}\BibitemShut {NoStop}%
\bibitem [{sup()}]{supplement}%
  \BibitemOpen
  \href@noop {} {\ }\bibinfo {note} {{Supplemental Material} mainly includes
  the electron injection in the laser-modulated wakefield, the temporal
  evolution of the spin precession frequency, parameterization of the sharp
  downward density jump, and the production of the electronic polarized gas
  target.}\BibitemShut {Stop}%
\bibitem [{\citenamefont {Mott-Smith}(1951)}]{mott1951}%
  \BibitemOpen
  \bibfield  {author} {\bibinfo {author} {\bibfnamefont {H.~M.}\ \bibnamefont
  {Mott-Smith}},\ }\bibfield  {title} {\bibinfo {title} {The {{Solution}} of
  the {{Boltzmann Equation}} for a {{Shock Wave}}},\ }\href
  {https://doi.org/10.1103/PhysRev.82.885} {\bibfield  {journal} {\bibinfo
  {journal} {Phys. Rev.}\ }\textbf {\bibinfo {volume} {82}},\ \bibinfo {pages}
  {885} (\bibinfo {year} {1951})}\BibitemShut {NoStop}%
\bibitem [{\citenamefont {Vieira}\ and\ \citenamefont {Mendon{\c
  c}a}(2014)}]{vieira2014}%
  \BibitemOpen
  \bibfield  {author} {\bibinfo {author} {\bibfnamefont {J.}~\bibnamefont
  {Vieira}}\ and\ \bibinfo {author} {\bibfnamefont {J.~T.}\ \bibnamefont
  {Mendon{\c c}a}},\ }\bibfield  {title} {\bibinfo {title} {Nonlinear {{Laser
  Driven Donut Wakefields}} for {{Positron}} and {{Electron Acceleration}}},\
  }\href {https://doi.org/10.1103/PhysRevLett.112.215001} {\bibfield  {journal}
  {\bibinfo  {journal} {Phys. Rev. Lett.}\ }\textbf {\bibinfo {volume} {112}},\
  \bibinfo {pages} {215001} (\bibinfo {year} {2014})}\BibitemShut {NoStop}%
\bibitem [{\citenamefont {Thomas}\ \emph {et~al.}(2020)\citenamefont {Thomas},
  \citenamefont {H{\"u}tzen}, \citenamefont {Lehrach}, \citenamefont {Pukhov},
  \citenamefont {Ji}, \citenamefont {Wu}, \citenamefont {Geng},\ and\
  \citenamefont {B{\"u}scher}}]{thomas2020}%
  \BibitemOpen
  \bibfield  {author} {\bibinfo {author} {\bibfnamefont {J.}~\bibnamefont
  {Thomas}}, \bibinfo {author} {\bibfnamefont {A.}~\bibnamefont {H{\"u}tzen}},
  \bibinfo {author} {\bibfnamefont {A.}~\bibnamefont {Lehrach}}, \bibinfo
  {author} {\bibfnamefont {A.}~\bibnamefont {Pukhov}}, \bibinfo {author}
  {\bibfnamefont {L.}~\bibnamefont {Ji}}, \bibinfo {author} {\bibfnamefont
  {Y.}~\bibnamefont {Wu}}, \bibinfo {author} {\bibfnamefont {X.}~\bibnamefont
  {Geng}},\ and\ \bibinfo {author} {\bibfnamefont {M.}~\bibnamefont
  {B{\"u}scher}},\ }\bibfield  {title} {\bibinfo {title} {Scaling laws for the
  depolarization time of relativistic particle beams in strong fields},\ }\href
  {https://doi.org/10.1103/PhysRevAccelBeams.23.064401} {\bibfield  {journal}
  {\bibinfo  {journal} {Phys. Rev. Accel. Beams}\ }\textbf {\bibinfo {volume}
  {23}},\ \bibinfo {pages} {064401} (\bibinfo {year} {2020})}\BibitemShut
  {NoStop}%
\bibitem [{\citenamefont {Mane}\ \emph {et~al.}(2005)\citenamefont {Mane},
  \citenamefont {Shatunov},\ and\ \citenamefont {Yokoya}}]{mane2005}%
  \BibitemOpen
  \bibfield  {author} {\bibinfo {author} {\bibfnamefont {S.~R.}\ \bibnamefont
  {Mane}}, \bibinfo {author} {\bibfnamefont {Y.~M.}\ \bibnamefont {Shatunov}},\
  and\ \bibinfo {author} {\bibfnamefont {K.}~\bibnamefont {Yokoya}},\
  }\bibfield  {title} {\bibinfo {title} {Spin-polarized charged particle beams
  in high-energy accelerators},\ }\href
  {https://doi.org/10.1088/0034-4885/68/9/R01} {\bibfield  {journal} {\bibinfo
  {journal} {Rep. Prog. Phys.}\ }\textbf {\bibinfo {volume} {68}},\ \bibinfo
  {pages} {1997} (\bibinfo {year} {2005})}\BibitemShut {NoStop}%
\bibitem [{\citenamefont {Arber}\ \emph {et~al.}(2015)\citenamefont {Arber},
  \citenamefont {Bennett}, \citenamefont {Brady}, \citenamefont
  {Lawrence-Douglas}, \citenamefont {Ramsay}, \citenamefont {Sircombe},
  \citenamefont {Gillies}, \citenamefont {Evans}, \citenamefont {Schmitz},
  \citenamefont {Bell},\ and\ \citenamefont {Ridgers}}]{arber2015}%
  \BibitemOpen
  \bibfield  {author} {\bibinfo {author} {\bibfnamefont {T.~D.}\ \bibnamefont
  {Arber}}, \bibinfo {author} {\bibfnamefont {K.}~\bibnamefont {Bennett}},
  \bibinfo {author} {\bibfnamefont {C.~S.}\ \bibnamefont {Brady}}, \bibinfo
  {author} {\bibfnamefont {A.}~\bibnamefont {Lawrence-Douglas}}, \bibinfo
  {author} {\bibfnamefont {M.~G.}\ \bibnamefont {Ramsay}}, \bibinfo {author}
  {\bibfnamefont {N.~J.}\ \bibnamefont {Sircombe}}, \bibinfo {author}
  {\bibfnamefont {P.}~\bibnamefont {Gillies}}, \bibinfo {author} {\bibfnamefont
  {R.~G.}\ \bibnamefont {Evans}}, \bibinfo {author} {\bibfnamefont
  {H.}~\bibnamefont {Schmitz}}, \bibinfo {author} {\bibfnamefont {A.~R.}\
  \bibnamefont {Bell}},\ and\ \bibinfo {author} {\bibfnamefont {C.~P.}\
  \bibnamefont {Ridgers}},\ }\bibfield  {title} {\bibinfo {title} {Contemporary
  particle-in-cell approach to laser-plasma modelling},\ }\href
  {https://doi.org/10.1088/0741-3335/57/11/113001} {\bibfield  {journal}
  {\bibinfo  {journal} {Plasma Physics and Controlled Fusion}\ }\textbf
  {\bibinfo {volume} {57}},\ \bibinfo {pages} {113001} (\bibinfo {year}
  {2015})}\BibitemShut {NoStop}%
\bibitem [{\citenamefont {Wan}\ \emph {et~al.}(2022)\citenamefont {Wan},
  \citenamefont {Wang}, \citenamefont {Zhao}, \citenamefont {Zhang},
  \citenamefont {Yu}, \citenamefont {Wang}, \citenamefont {Yan}, \citenamefont
  {Zhao}, \citenamefont {Hatsagortsyan}, \citenamefont {Keitel}, \citenamefont
  {Bulanov},\ and\ \citenamefont {Li}}]{wan2022}%
  \BibitemOpen
  \bibfield  {author} {\bibinfo {author} {\bibfnamefont {F.}~\bibnamefont
  {Wan}}, \bibinfo {author} {\bibfnamefont {W.-Q.}\ \bibnamefont {Wang}},
  \bibinfo {author} {\bibfnamefont {Q.}~\bibnamefont {Zhao}}, \bibinfo {author}
  {\bibfnamefont {H.}~\bibnamefont {Zhang}}, \bibinfo {author} {\bibfnamefont
  {T.-P.}\ \bibnamefont {Yu}}, \bibinfo {author} {\bibfnamefont {W.-M.}\
  \bibnamefont {Wang}}, \bibinfo {author} {\bibfnamefont {W.-C.}\ \bibnamefont
  {Yan}}, \bibinfo {author} {\bibfnamefont {Y.-T.}\ \bibnamefont {Zhao}},
  \bibinfo {author} {\bibfnamefont {K.~Z.}\ \bibnamefont {Hatsagortsyan}},
  \bibinfo {author} {\bibfnamefont {C.~H.}\ \bibnamefont {Keitel}}, \bibinfo
  {author} {\bibfnamefont {S.~V.}\ \bibnamefont {Bulanov}},\ and\ \bibinfo
  {author} {\bibfnamefont {J.-X.}\ \bibnamefont {Li}},\ }\bibfield  {title}
  {\bibinfo {title} {Quasimonoenergetic {{Proton Acceleration}} via {{Quantum
  Radiative Compression}}},\ }\href
  {https://doi.org/10.1103/PhysRevApplied.17.024049} {\bibfield  {journal}
  {\bibinfo  {journal} {Phys. Rev. Applied}\ }\textbf {\bibinfo {volume}
  {17}},\ \bibinfo {pages} {024049} (\bibinfo {year} {2022})}\BibitemShut
  {NoStop}%
\bibitem [{\citenamefont {Carbajo}\ \emph {et~al.}(2014)\citenamefont
  {Carbajo}, \citenamefont {Granados}, \citenamefont {Schimpf}, \citenamefont
  {Sell}, \citenamefont {Hong}, \citenamefont {Moses},\ and\ \citenamefont
  {Kaertner}}]{carbajo2014}%
  \BibitemOpen
  \bibfield  {author} {\bibinfo {author} {\bibfnamefont {S.}~\bibnamefont
  {Carbajo}}, \bibinfo {author} {\bibfnamefont {E.}~\bibnamefont {Granados}},
  \bibinfo {author} {\bibfnamefont {D.}~\bibnamefont {Schimpf}}, \bibinfo
  {author} {\bibfnamefont {A.}~\bibnamefont {Sell}}, \bibinfo {author}
  {\bibfnamefont {K.-H.}\ \bibnamefont {Hong}}, \bibinfo {author}
  {\bibfnamefont {J.}~\bibnamefont {Moses}},\ and\ \bibinfo {author}
  {\bibfnamefont {F.~X.}\ \bibnamefont {Kaertner}},\ }\bibfield  {title}
  {\bibinfo {title} {Efficient generation of ultra-intense few-cycle radially
  polarized laser pulses},\ }\href {https://doi.org/10.1364/OL.39.002487}
  {\bibfield  {journal} {\bibinfo  {journal} {Opt. Lett.}\ }\textbf {\bibinfo
  {volume} {39}},\ \bibinfo {pages} {2487} (\bibinfo {year}
  {2014})}\BibitemShut {NoStop}%
\bibitem [{\citenamefont {Kong}\ \emph {et~al.}(2019)\citenamefont {Kong},
  \citenamefont {Larocque}, \citenamefont {Karimi}, \citenamefont {Corkum},\
  and\ \citenamefont {Zhang}}]{kong2019}%
  \BibitemOpen
  \bibfield  {author} {\bibinfo {author} {\bibfnamefont {F.}~\bibnamefont
  {Kong}}, \bibinfo {author} {\bibfnamefont {H.}~\bibnamefont {Larocque}},
  \bibinfo {author} {\bibfnamefont {E.}~\bibnamefont {Karimi}}, \bibinfo
  {author} {\bibfnamefont {P.~B.}\ \bibnamefont {Corkum}},\ and\ \bibinfo
  {author} {\bibfnamefont {C.}~\bibnamefont {Zhang}},\ }\bibfield  {title}
  {\bibinfo {title} {Generating few-cycle radially polarized pulses},\ }\href
  {https://doi.org/10.1364/OPTICA.6.000160} {\bibfield  {journal} {\bibinfo
  {journal} {Optica}\ }\textbf {\bibinfo {volume} {6}},\ \bibinfo {pages} {160}
  (\bibinfo {year} {2019})}\BibitemShut {NoStop}%
\bibitem [{\citenamefont {Zhong}\ \emph {et~al.}(2021)\citenamefont {Zhong},
  \citenamefont {Liang}, \citenamefont {Dai}, \citenamefont {Huang},
  \citenamefont {Hu}, \citenamefont {Xu},\ and\ \citenamefont
  {Qian}}]{zhong2021}%
  \BibitemOpen
  \bibfield  {author} {\bibinfo {author} {\bibfnamefont {H.}~\bibnamefont
  {Zhong}}, \bibinfo {author} {\bibfnamefont {C.}~\bibnamefont {Liang}},
  \bibinfo {author} {\bibfnamefont {S.}~\bibnamefont {Dai}}, \bibinfo {author}
  {\bibfnamefont {J.}~\bibnamefont {Huang}}, \bibinfo {author} {\bibfnamefont
  {S.}~\bibnamefont {Hu}}, \bibinfo {author} {\bibfnamefont {C.}~\bibnamefont
  {Xu}},\ and\ \bibinfo {author} {\bibfnamefont {L.}~\bibnamefont {Qian}},\
  }\bibfield  {title} {\bibinfo {title} {Polarization-insensitive, high-gain
  parametric amplification of radially polarized femtosecond pulses},\ }\href
  {https://doi.org/10.1364/OPTICA.413328} {\bibfield  {journal} {\bibinfo
  {journal} {Optica}\ }\textbf {\bibinfo {volume} {8}},\ \bibinfo {pages} {62}
  (\bibinfo {year} {2021})}\BibitemShut {NoStop}%
\end{thebibliography}%

\end{document}